\documentclass[aps,prl,twocolumn,groupedaddress,showpacs,floatfix,superscriptaddress,longbibliography]{revtex4-2}
\usepackage[plainpages=false,pdfpagelabels,colorlinks=true,linkcolor=red,urlcolor=blue,citecolor=blue,pdftitle={Title},pdfauthor={},pdfdisplaydoctitle=true,pdfduplex=DuplexFlipLongEdge]{hyperref}
\usepackage{siunitx}
\usepackage{mhchem}
\usepackage{epsfig}
\usepackage{graphicx}
\usepackage[utf8]{inputenc}
\usepackage{amsmath,amssymb}
\usepackage[dvipsnames]{xcolor}

\usepackage[normalem]{ulem}

\usepackage{tabularx}
\usepackage{multirow}
\usepackage{multibib}
\newcites{SI}{References}

\makeatletter
\edef\MainNotesBib{\jobname Notes}
\let\oldbibliographySI\bibliographySI
\renewcommand{\bibliographySI}[1]{%
	\begingroup
	\def\pre@bibdata{\MainNotesBib}%
	\oldbibliographySI{#1}%
	\endgroup
}
\makeatother

\usepackage{array}
\usepackage{float}
\begin{document}
	

\title{Vacancy-Driven Electronic Reconstruction in Monolayer PtSe$_2$: Formation Thermodynamics and Charge States}

\author{Xinwen Gai}
\affiliation{School of Physics and Astronomy, Beijing Normal University, and Key Laboratory of Multiscale Spin Physics (Beijing Normal University), Ministry of Education, Beijing 100875, China} 

\author{Jingang Wang}
\affiliation{College of Science, Liaoning Petrochemical University, Fushun 113001, China}

\author{Tianxing Ma}
\email{txma@bnu.edu.cn}
\affiliation{School of Physics and Astronomy, Beijing Normal University, and Key Laboratory of Multiscale Spin Physics (Beijing Normal University), Ministry of Education, Beijing 100875, China}

\begin{abstract}
Layered transition metal dichalcogenides are an important platform for two-dimensional materials, where the inevitable intrinsic defects provide new degrees of freedom for tuning their physical properties. Based on first-principles calculations, this work systematically investigates the formation energies, charge states, and electronic structural characteristics of V$_{\mathrm{Pt}}$, V$_{\mathrm{Se}}$, and composite vacancies in monolayer PtSe$_2$. The results indicate that while vacancy formation energies are highly sensitive to chemical potentials, the V$_{\mathrm{Se}}$ structure consistently exhibits the lowest formation energy. The charge defect calculation reveals the stable charge state intervals of V$_{\mathrm{Se}}$ and V$_{\mathrm{Pt}}$ as a function of the Fermi level, thus describing the evolution of the charge states of intrinsic vacancies at different electronic chemical potentials.
Climbing Image Nudged Elastic Band calculations reveal high migration barriers for V$_{\mathrm{Se}}$ and V$_{\mathrm{Pt}}$, indicating strongly hindered vacancy diffusion at room temperature, while short Ab Initio Molecular Dynamics simulations confirm the absence of immediate structural collapse within the simulated time window.
Optical property calculations indicate that point defects significantly alter the dielectric response of monolayer PtSe$_2$ and generate new low-energy absorption channels associated with in-gap defect states. These findings provide new insights into defect-mediated electronic and optical property modulation in monolayer PtSe$_2$, offering guidance for its potential device design.
\end{abstract}
\maketitle


\section{Introduction}
Two-dimensional (2D) transition metal dichalcogenides (TMDs) represent an important branch of layered van der Waals materials. Owing to their quantum-size effects, thickness-dependent band structures, and excellent mechanical flexibility, they have been widely regarded as promising candidates for next-generation optoelectronic and spintronic devices~\cite{1,2,3}. Over the past decade, 2D TMDs have emerged as a key platform for studying low-dimensional quantum physics, including topological states~\cite{4,5}, valley physics~\cite{6,7}, and defect-induced magnetism~\cite{8,9}. Nevertheless, defects in real materials are almost unavoidable, including vacancies, interstitials, and antisite defects~\cite{10,11}. Such defects can introduce localized states within the band gap, modulate the Fermi level and carrier concentration, and may even serve as key factors in inducing magnetism or enhancing catalytic activity~\cite{12,13,14,15}. On the other hand, defects can also be intentionally introduced through irradiation, electron-beam exposure, or vacuum annealing to tailor the physical properties of 2D materials~\cite{16,17,18}. Therefore, from a theoretical perspective, understanding the formation mechanisms of defects in 2D TMDs and their influence on electronic properties is of significant scientific importance.

Among the various TMDs, PtSe$_2$ has attracted considerable attention as a representative 1T-phase layered compound due to its high chemical stability and tunable electronic band structure~\cite{19,20}. Unlike the hexagonal MoS$_2$ and WS$_2$ in the 2H phase, Pt atoms in PtSe$_2$ reside in an octahedral coordination environment, forming PtSe$_6$ structural units that assemble into a 2D network, resulting in electronic characteristics that differ markedly from those of conventional 2H-phase TMDs~\cite{21}. In recent years, PtSe$_2$ has exhibited excellent performance in photodetection~\cite{22,23}, gas sensing~\cite{24,25}, and catalytic reactions~\cite{26,27}, thereby stimulating extensive research interest. PtSe$_2$ is also known to host Se vacancies, Pt vacancies, and their composite defect configurations~\cite{28}, which introduce in-gap electronic states, modulate the band dispersion, and may even induce localized magnetism. 

Gao et al. systematically investigated the structures, stability, electronic properties, and migration behavior of neutral Se and Pt single and double vacancies in monolayer PtSe$_2$~\cite{29}. It showed that Se and Pt vacancies can significantly modify the electronic structure of monolayer PtSe$_2$, and that Pt single and double vacancies can induce pronounced spin polarization. Zheng et al. further combined STM/STS measurements with first-principles calculations to identify various intrinsic point defects in ultrathin 1T-PtSe$_2$, including Se vacancies, Pt vacancies, and Se$_{\mathrm{Pt}}$ antisite defects~\cite{28}. Subsequently, Avsar et al. reported defect-induced magnetic ordering in atomically thin semiconducting PtSe$_2$ and identified Pt vacancies as the key origin of the induced magnetism~\cite{30}. The catalytic role of PtSe$_2$ defects has also attracted attention. Chang et al. reported the bifunctional HER/OER catalytic performance of Se-vacancy-engineered PtSe$_2$~\cite{27,31}, showing that Se vacancies can introduce localized defect states and charge trapping, thereby enhancing adsorption and reaction activity. These findings indicate that intrinsic vacancies in PtSe$_2$ not only affect electronic transport and magnetism~\cite{30,32}, but may also regulate chemical adsorption and catalytic reactions. A detailed comparison between the present work and representative previous PtSe$_2$ defect studies is provided in Table~S1.

However, systematic investigations of intrinsic charged defects in monolayer PtSe$_2$ remain relatively limited, and their formation thermodynamics, stable charge state ranges, and possible kinetic behaviors have yet to be fully elucidated. In addition, the actual influence of defects is determined not only by their formation energies, but also by their migration kinetics and finite-temperature structural stability~\cite{29,33}. Therefore, it is necessary to establish a unified theoretical framework that simultaneously covers defect formation thermodynamics, charge-state stability, electronic structure reconstruction, kinetic stability, and reaction activity.

Based on this background, this work employs first-principles calculations to systematically investigate the formation thermodynamics, charge state stability, electronic structure reconstruction, and finite-temperature kinetic behavior of different intrinsic vacancies in monolayer PtSe$_2$. We further analyze the effects of these defects on the optical properties and hydrogen adsorption thermodynamics, aiming to reveal how vacancies regulate the local reaction activity of PtSe$_2$. The present results provide a unified picture for understanding defect physics in PtSe$_2$ and offer a theoretical basis for defect engineering and catalytic applications in two-dimensional TMDs.

\section{Computational Methods}
All first-principles calculations in this work were carried out within the framework of density functional theory (DFT)~\cite{34,35} using the Vienna \textit{Ab initio} Simulation Package (VASP 6.3.2)~\cite{36}. The ion-electron interactions were described using the projector augmented wave (PAW) method. Defect models were constructed based on a monolayer PtSe$_2$ structure, employing a 6$\times$6$\times$1 supercell together with a 20~\AA\ vacuum layer along the $z$ direction, which was verified to be sufficient to suppress spurious interactions between periodic images. Therefore, dipole corrections were not applied. Structural optimizations were performed using the Perdew--Burke--Ernzerhof (PBE) functional within the generalized gradient approximation (GGA) at the scalar-relativistic level~\cite{37,38}. Given that spin--orbit coupling (SOC) typically has a significant impact on electronic properties, SOC was explicitly included in all subsequent VASP single-point calculations to ensure an accurate and consistent description of the electronic structures, defect states, magnetic moments, charged defects, and optical properties. Brillouin-zone integrations were carried out using $\Gamma$-centered $16\times16\times1$ and $2\times2\times1$ $k$-point meshes for the primitive cell and supercell, respectively~\cite{39}. The plane-wave kinetic energy cutoff was set to 500~eV. The convergence criteria for structural relaxation were 10$^{-6}$~eV for the total energy and 0.02~eV/\AA\ for the atomic forces. Van der Waals interactions were included using the Grimme DFT-D3 correction scheme to ensure a reliable description of the layered system~\cite{40}.
Electronic properties were analyzed through band structure and density-of-states (DOS) calculations, with the band structure computed along the high-symmetry path M--K--$\Gamma$--M of the hexagonal PtSe$_2$ lattice. Defect diffusion pathways were initially optimized using the climbing image nudged elastic band (CI-NEB) method~\cite{41,42} at the scalar-relativistic level. To accurately capture the energy landscape, the migration barriers were subsequently refined by performing single-point energy calculations with SOC on the converged NEB images.
To evaluate the local charge redistribution, the Charge Model 5 (CM5) analysis~\cite{43} was employed. Both the CM5 charge calculations and the \textit{ab initio} molecular dynamics (AIMD) simulations were performed using the CP2K 2025.2 software package~\cite{44,45}, which utilizes localized Gaussian basis sets that are highly robust for void-containing structures. These calculations employed GTH-PBE pseudopotentials and the DZVP-MOLOPT-SR-GTH basis set, with the plane-wave density cutoff and relative cutoff set to 500 Ry and 55 Ry, respectively. The simulations were conducted in the canonical ensemble, namely the NVT ensemble, with the CSVR thermostat used to maintain the system temperature at 300 K. The time step was set to 1 fs, and the total simulation duration was 10 ps.

\section{Results and Discussion}
\begin{figure*}[htbp]
	\centering
	\includegraphics[width=0.95\textwidth]{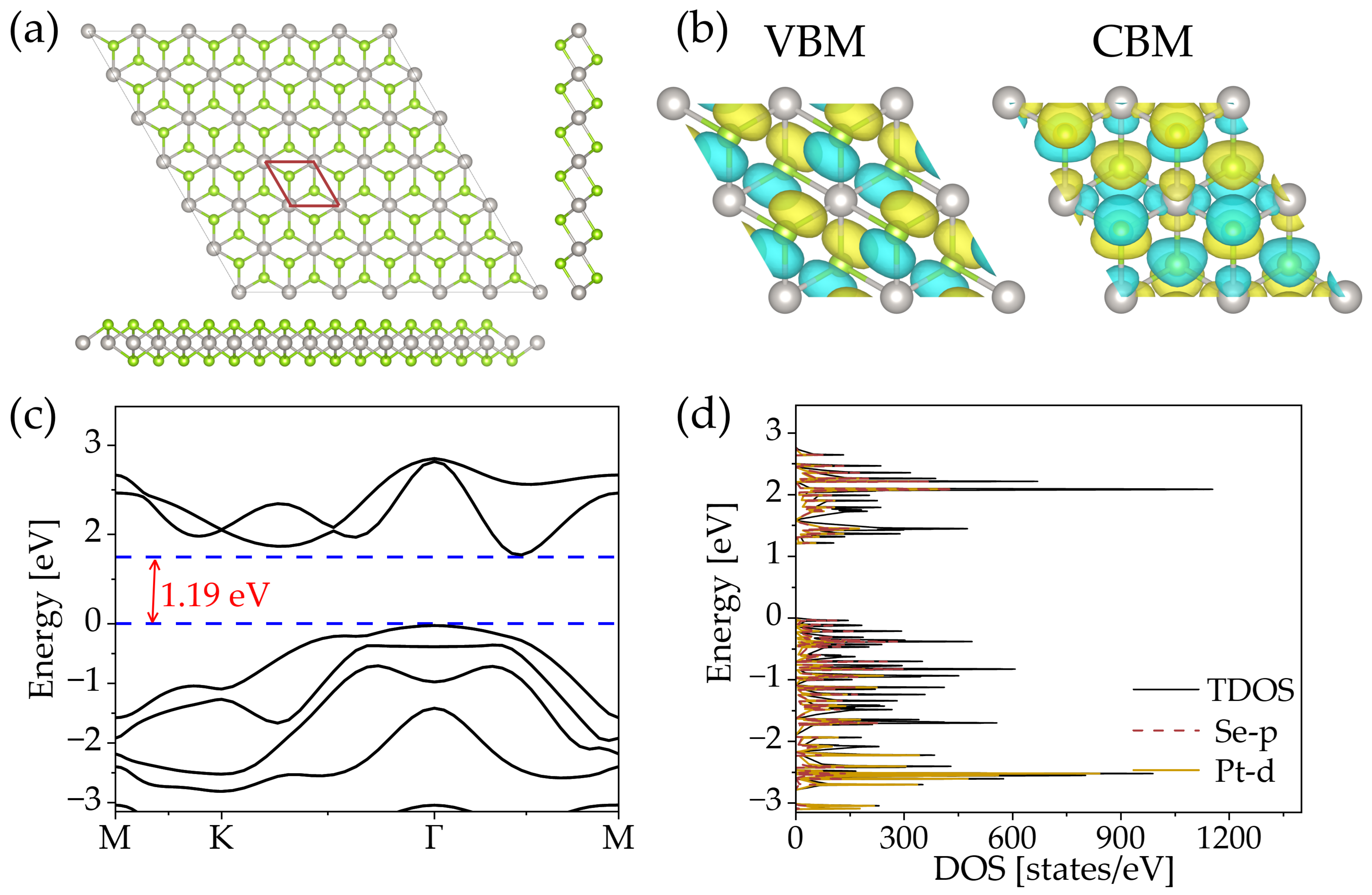}
	\caption{(a) Schematic crystal structure of monolayer PtSe$_2$ (top and side views). 
		Gray and green spheres represent Pt and Se atoms, respectively. 
		(b) Spatial charge density distributions of the VBM and CBM, with isosurface values of 0.0002~e/\AA$^3$ and 0.0007~e/\AA$^3$, respectively; yellow and blue regions denote positive and negative phases. 
		(c) Electronic band structure of monolayer PtSe$_2$, with the Fermi level set as the energy reference at 0~eV. 
		(d) Total DOS (TDOS) and projected DOS (PDOS).}
	\label{fig:fig1}
\end{figure*}
The crystal structure of monolayer PtSe$_2$ is shown in Fig.~\ref{fig:fig1}(a). The model adopts a 6$\times$6$\times$1 supercell containing 108 atoms, in which Pt atoms occupy the middle layer and are octahedrally coordinated by six Se atoms from the upper and lower layers; each Se atom is bonded to three Pt atoms, forming a typical 1T-phase layered structure. The primitive cell of monolayer PtSe$_2$ contains three atoms (one Pt and two Se), as indicated by the red parallelogram in Fig.~\ref{fig:fig1}(a). Geometry optimization yields an in-plane lattice constant of $a = 3.72$~\AA, a Pt--Se bond length of approximately 2.53~\AA, and a monolayer thickness of about 2.62~\AA, defined as the vertical distance between the upper and lower Se atomic layers. These structural parameters are in good agreement with previously reported experimental and theoretical values~\cite{46}.

As shown in Fig.~\ref{fig:fig1}(b), the charge density at the valence band maximum (VBM) is mainly distributed around the Se atoms in the upper and lower layers, exhibiting pronounced $p_x/p_y$ orbital characteristics, while no electron density is found at the Pt sites. This indicates that the VBM is primarily derived from Se--$p$ states. In contrast, the charge density at the conduction band minimum (CBM) is localized around the Pt atoms and displays the typical directional features of $d$ orbitals; at the same time, a certain amount of electron density is also observed around the neighboring Se atoms, matching the spatial distribution of the Pt--$d$ states, suggesting that the CBM originates from the hybridization between Pt--$d$ and Se--$p$ orbitals.

To further elucidate the intrinsic electronic structure of monolayer $\text{PtSe}_2$, the band structure calculated using the PBE+SOC method is presented in Fig.~\ref{fig:fig1}(c). The band profile indicates that the system is a typical indirect band gap semiconductor. Calculations performed at different theoretical levels yield band gaps of 1.40 eV, 1.19 eV, and 1.95 eV for the PBE, PBE+SOC, and HSE06 methods, respectively, which are highly consistent with theoretical predictions in the existing literature~\cite{9,29,47,48,49,50,51}. In particular, the HSE06 result is in excellent agreement with the experimentally measured band gap of approximately 2.0 eV. The band structures calculated using the GGA-PBE and HSE06 methods are shown in Fig.~S1 of the Supporting Information. Furthermore, based on the PDOS shown in Fig.~\ref{fig:fig1}(d), it is evident that the VBM is primarily contributed by the Se--$p$ states, whereas the CBM is jointly dominated by the Pt--$d$ and Se--$p$ states. This demonstrates that the electronic properties of monolayer $\text{PtSe}_2$ exhibit pronounced orbital hybridization characteristics, providing an important physical reference for subsequent studies on the modulation of defect states.

Point defects are almost unavoidable during the practical synthesis of 2D materials, particularly vacancy-type defects. With the advancement of experimental techniques, methods such as electron irradiation and ion implantation have further enabled the controlled introduction of defects at the atomic scale~\cite{52,53}. Among various TMDs, chalcogen vacancies are generally considered to be the most common intrinsic defects. However, owing to the unique 1T-phase octahedral coordination environment of PtSe$_2$, its physical properties can be particularly sensitive to vacancy formation. Such vacancy defects not only introduce localized states within the band gap but can also significantly modulate the optical, electronic, and even magnetic properties of the material, thereby offering new opportunities for controllable property engineering. Therefore, systematically investigating representative vacancy defects in monolayer PtSe$_2$ is of both scientific importance and practical relevance.
\begin{figure}[htbp]
	\centering
	\includegraphics[width=0.45\textwidth]{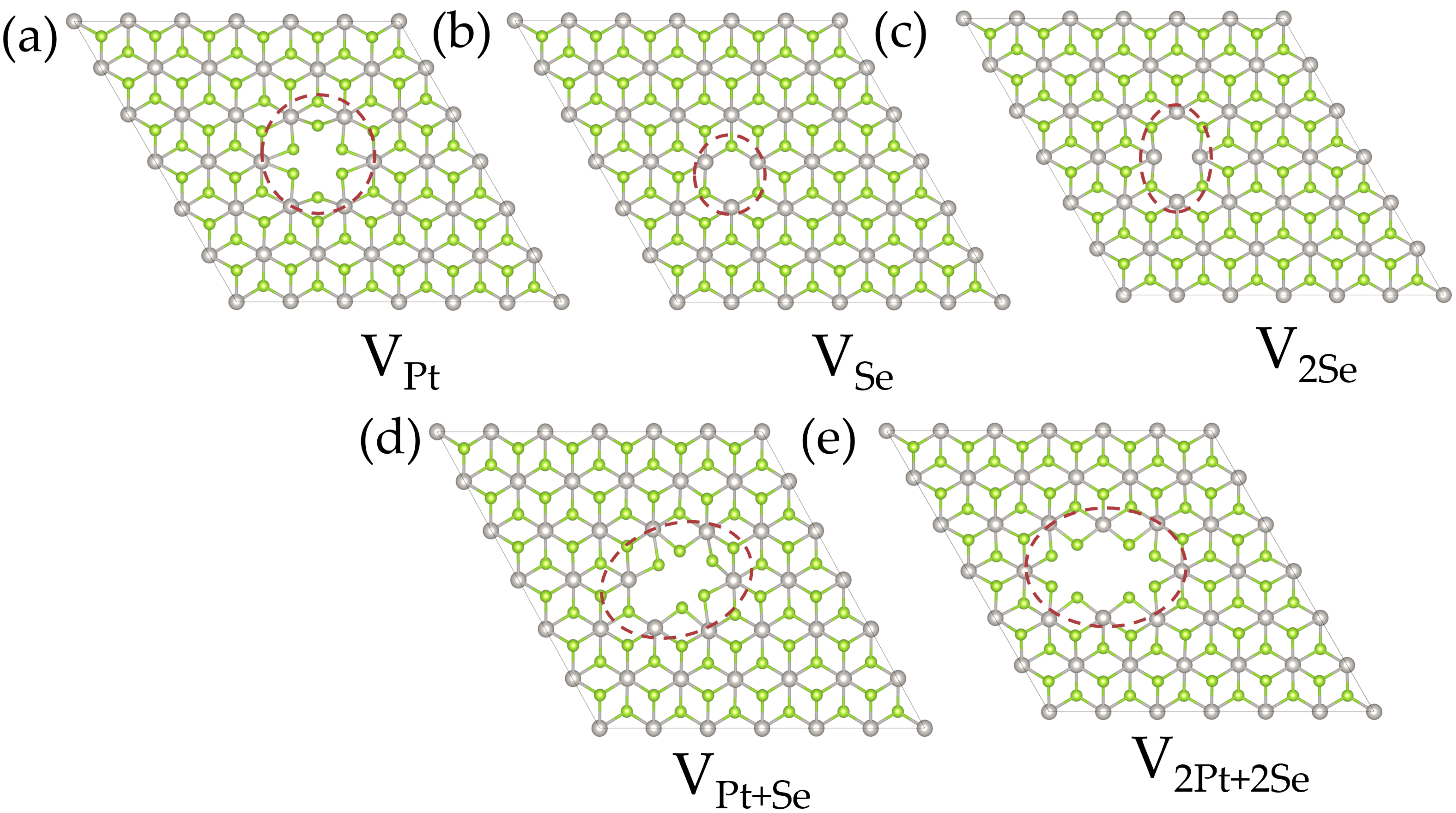}
	\caption{Schematic structures of representative intrinsic defects in monolayer PtSe$_2$ with (a) V$_{\mathrm{Pt}}$, (b) V$_{\mathrm{Se}}$, (c) V$_{2\mathrm{Se}}$, (d) V$_{\mathrm{Pt+Se}}$, and (e) V$_{2\mathrm{Pt}+2\mathrm{Se}}$. Defect sites are highlighted by red dashed circles.}
	\label{fig:fig2}
\end{figure}

Based on the established crystal and electronic structures of pristine monolayer PtSe$_2$, several representative intrinsic vacancy defects were constructed, as shown in  Fig.~\ref{fig:fig2}. All calculations were performed using the 6$\times$6$\times$1 supercell model, ensuring that the minimum separation between periodic defect images exceeds 10~\AA, thus effectively eliminating spurious interactions and approximating isolated defects. The considered defect configurations include: a Pt monovacancy (V$_{\mathrm{Pt}}$), a Se monovacancy (V$_{\mathrm{Se}}$), a Se divacancy formed by removing two adjacent Se atoms (V$_{\mathrm{2Se}}$), a composite vacancy involving a Pt atom and its neighboring Se atom (V$_{\mathrm{Pt+Se}}$), and a larger multi-atom composite vacancy (V$_{\mathrm{2Pt+2Se}}$). All defective structures were fully relaxed to their equilibrium geometries. The defect sites are highlighted with red dashed circles for clarity.

The thermodynamic stability of different vacancy configurations is evaluated by calculating their defect formation energies, and the defect formation energy is defined as
\begin{equation}
	E_{\mathrm{f}} = E_{\mathrm{def}} - E_{\mathrm{pristine}} + \sum_i n_i \mu_i,
	\label{eq:defect_formation}
\end{equation}
where $E_{\mathrm{def}}$ and $E_{\mathrm{pristine}}$ denote the total energies of the defective and pristine monolayer PtSe$_2$, respectively; $n_i$ is the number of atoms of species $i$ removed from the supercell when creating the defect, and $\mu_i$ is the corresponding chemical potential. The chemical potentials depend on the growth environment and are constrained by thermodynamic equilibrium. In particular, their values cannot exceed those of the elements in their most stable bulk phases. For defect-free monolayer PtSe$_2$, the chemical potentials satisfy
\begin{equation}
	\mu_{\mathrm{PtSe_2}} = \mu_{\mathrm{Pt}} + 2\mu_{\mathrm{Se}}
	= \mu_{\mathrm{Pt}}^{\mathrm{bulk}} + 2\mu_{\mathrm{Se}}^{\mathrm{bulk}} - \Delta H_{\mathrm{f}}(\mathrm{PtSe_2}),
\end{equation}
where $\mu_{\mathrm{Pt}}^{\mathrm{bulk}}$ and $\mu_{\mathrm{Se}}^{\mathrm{bulk}}$ are the chemical potentials of Pt and Se in their most stable bulk phases (face-centered-cubic Pt, space group Fm$\bar{3}$m, and hexagonal Se, space group P3$_1$21), and $\Delta H_{\mathrm{f}}(\mathrm{PtSe_2})$ is the formation enthalpy of monolayer PtSe$_2$. This leads to the following bounds for the chemical potentials:
\begin{equation}
	\mu_{\mathrm{Pt}}^{\mathrm{bulk}} - \Delta H_{\mathrm{f}}(\mathrm{PtSe_2})
	\;\le\;
	\mu_{\mathrm{Pt}}
	\;\le\;
	\mu_{\mathrm{Pt}}^{\mathrm{bulk}},
\end{equation}
\begin{equation}
	\mu_{\mathrm{Se}}^{\mathrm{bulk}} - \frac{1}{2}\Delta H_{\mathrm{f}}(\mathrm{PtSe_2})
	\;\le\;
	\mu_{\mathrm{Se}}
	\;\le\;
	\mu_{\mathrm{Se}}^{\mathrm{bulk}}.
\end{equation}

Under the elemental-reservoir Pt-rich limit, \(\mu_{\mathrm{Pt}}=\mu_{\mathrm{Pt}}^{\mathrm{bulk}}\), where the chemical potential of Pt is \(-6.483~\mathrm{eV}\), and the corresponding \(\mu_{\mathrm{Se}}\) is \(-4.125~\mathrm{eV}\). Under Se-rich conditions, \(\mu_{\mathrm{Se}}=\mu_{\mathrm{Se}}^{\mathrm{bulk}}\), where \(\mu_{\mathrm{Se}}\) is \(-3.436~\mathrm{eV}\), and the corresponding \(\mu_{\mathrm{Pt}}\) is \(-7.860~\mathrm{eV}\). After further considering the competing Pt--Se phase Pt$_5$Se$_4$, the chemical potentials should satisfy \(5\mu_{\mathrm{Pt}}+4\mu_{\mathrm{Se}}\leq \mu_{\mathrm{Pt}_5\mathrm{Se}_4}\) to avoid the precipitation of Pt$_5$Se$_4$. The calculated energy of Pt$_5$Se$_4$ is \(-49.03605\) eV per formula unit. By substituting the stability condition of PtSe$_2$ into this constraint, \(\mu_{\mathrm{Se}}\geq -4.105\) eV is obtained, corresponding to \(\mu_{\mathrm{Pt}}=-6.523\) eV. Therefore, the Pt$_5$Se$_4$ competing phase only constrains the Pt-rich/Se-poor limit and revises it from \(\mu_{\mathrm{Se}}=-4.125\) eV to \(\mu_{\mathrm{Se}}=-4.105\) eV.

\begin{figure}[t]  
	\centering
	\includegraphics[width=0.45\textwidth]{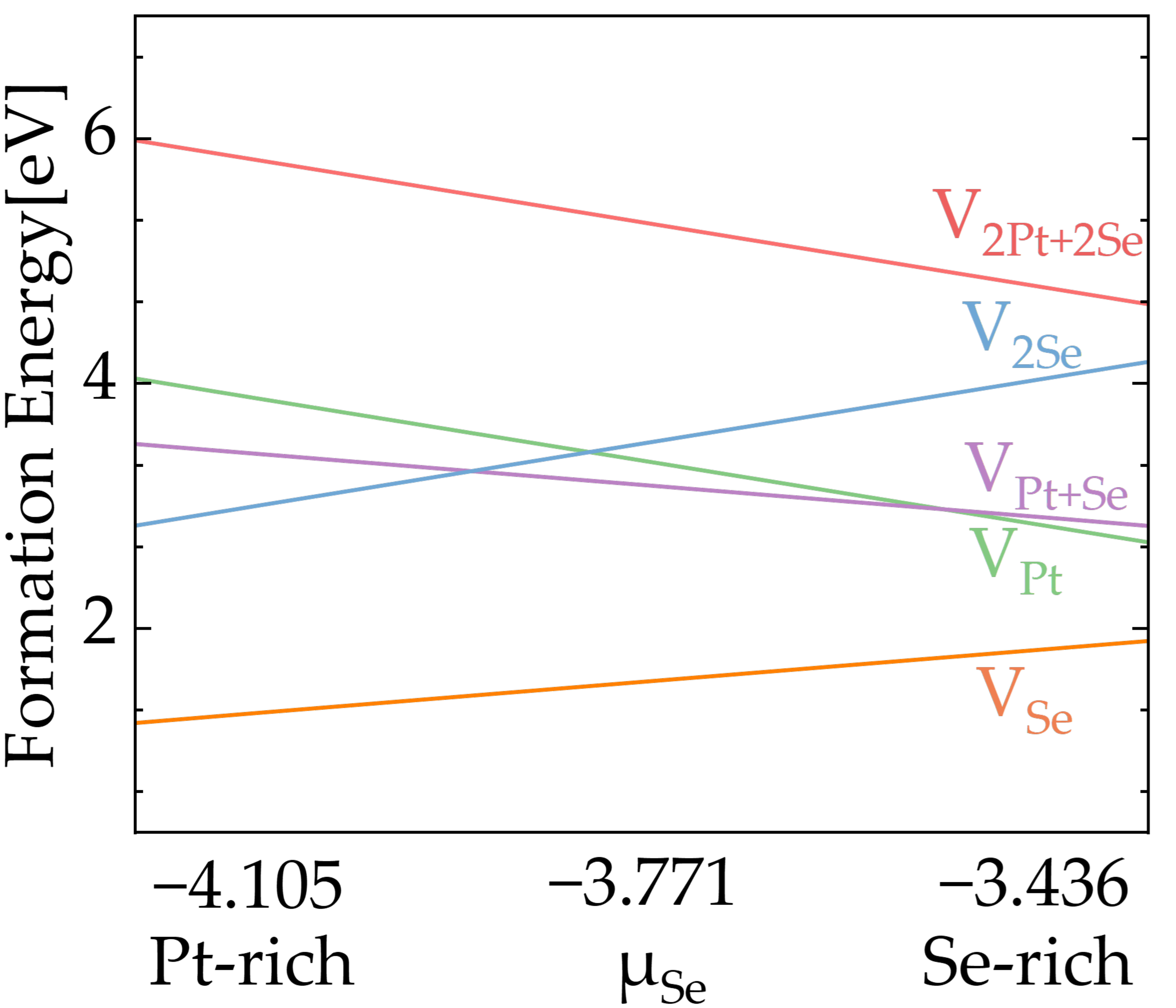}
	\caption{Formation energies of different intrinsic defects in monolayer PtSe$_2$ as a function of $
		\mu_\text{Se}$.}
	\label{fig:fig3}
\end{figure}
\begin{figure*}[htbp]
	\centering
	\includegraphics[width=0.86\textwidth]{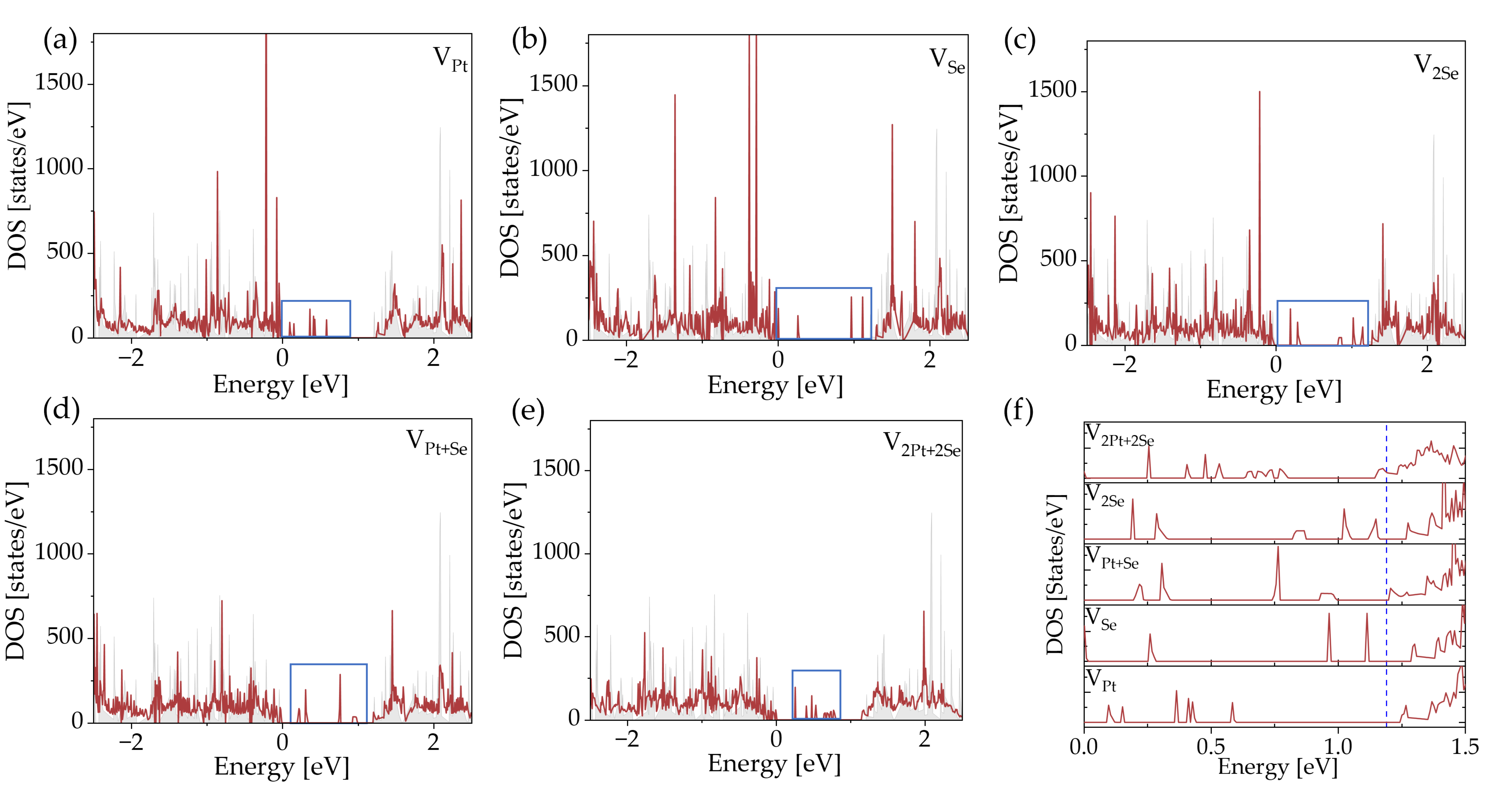}
	\caption{DOS for defected monolayer PtSe$_2$: (a) V$_{\mathrm{Pt}}$, (b) V$_{\mathrm{Se}}$, (c) V$_{\mathrm{2Se}}$, (d) V$_{\mathrm{Pt+Se}}$, and (e) V$_{\mathrm{2Pt+2Se}}$ (red curves), compared with the defect-free monolayer PtSe$_2$ (gray fill). The Fermi level is set to 0 eV. (f) Comparative plot for different defects.}
	\label{fig:fig4}
\end{figure*}

{\color{red}\begin{table}[b]
		\caption{\label{TabA1}Formation energies of vacancies (eV) in a PtSe$_2$ monolayer.}
		\renewcommand{\arraystretch}{1.5}
		\begin{ruledtabular}
			\begin{tabular}{lccccc}
				site   &  V$_{\mathrm{Se}}$      &  V$_{\mathrm{Pt}}$     & V$_{\mathrm{2Se}}$      & V$_{\mathrm{Pt+Se}}$    & V$_{\mathrm{2Pt+2Se}}$   \\
				\colrule
				Pt-rich & 1.23 & 4.04 & 2.84 & 3.50 & 5.98 \\
				Se-rich & 1.89 & 2.70 & 4.17 & 2.84 & 4.65 \\
			\end{tabular}
		\end{ruledtabular}
\end{table}}

Fig.~\ref{fig:fig3} shows the trend of defect formation energies for different vacancy types as a function of the Se chemical potential, with the corresponding values listed in TABLE~\ref{TabA1}, representing the Pt-rich and Se-rich limits, respectively. Overall, the formation energies of all defects are positive, indicating that their thermodynamic stabilities are lower than that of the perfect PtSe$_2$ crystal. However, different types of defects exhibit significant differences under different chemical environments. Throughout the entire allowable range of chemical potential, the mono-vacancy V$_{\mathrm{Se}}$ exhibits the lowest formation energy, indicating it is the most thermodynamically stable defect. 

Under Pt-rich conditions, the formation energy of V$_{\mathrm{Se}}$ is as low as  1.23~eV, indicating that it is the most favorable vacancy under Pt-rich conditions. 
In this regime, the divacancy V$_{\mathrm{2Se}}$ also possesses a relatively low formation energy (2.84~eV), representing the second most stable defect configuration. 
In contrast, V$_{\mathrm{Pt}}$ and complex defects (V$_{\mathrm{Pt+Se}}$, V$_{\mathrm{2Pt+2Se}}$) exhibit significantly higher formation energies under Pt-rich conditions, rendering them energetically unfavorable.
As the chemical potential shifts toward the Se-rich limit, the formation energy of V$_{\mathrm{Pt}}$ decreases significantly to 2.70~eV, indicating that a Se-rich environment facilitates the generation of Pt vacancies. 
Although the formation energy of the large complex vacancy V$_{\mathrm{2Pt+2Se}}$ remains consistently higher than that of other defects, it displays a downward trend under Se-rich conditions. 
While V$_{\mathrm{2Pt+2Se}}$ is difficult to form under thermal equilibrium, it may still be induced by non-equilibrium processes such as external irradiation, ion bombardment, or high-temperature annealing.

Furthermore, the sum of the formation energies of V$_{\mathrm{Pt}}$ and V$_{\mathrm{Se}}$ is higher than the formation energy of the composite defect V$_{\mathrm{Pt+Se}}$, indicating that the energy of the system decreases when these two vacancies combine, suggesting an attractive interaction between them. Therefore, V$_{\mathrm{Pt+Se}}$ is slightly more stable than the two isolated vacancies. Similarly, for the Se vacancy system, the sum of the formation energies of two individual V$_{\mathrm{Se}}$ is lower than the formation energy of V$_{2\mathrm{Se}}$, implying that there is a repulsive interaction between the two Se vacancies. Their aggregation into a divacancy does not lower the system's energy.

\begin{figure*}[htbp]
	\centering
	\includegraphics[width=0.86\textwidth]{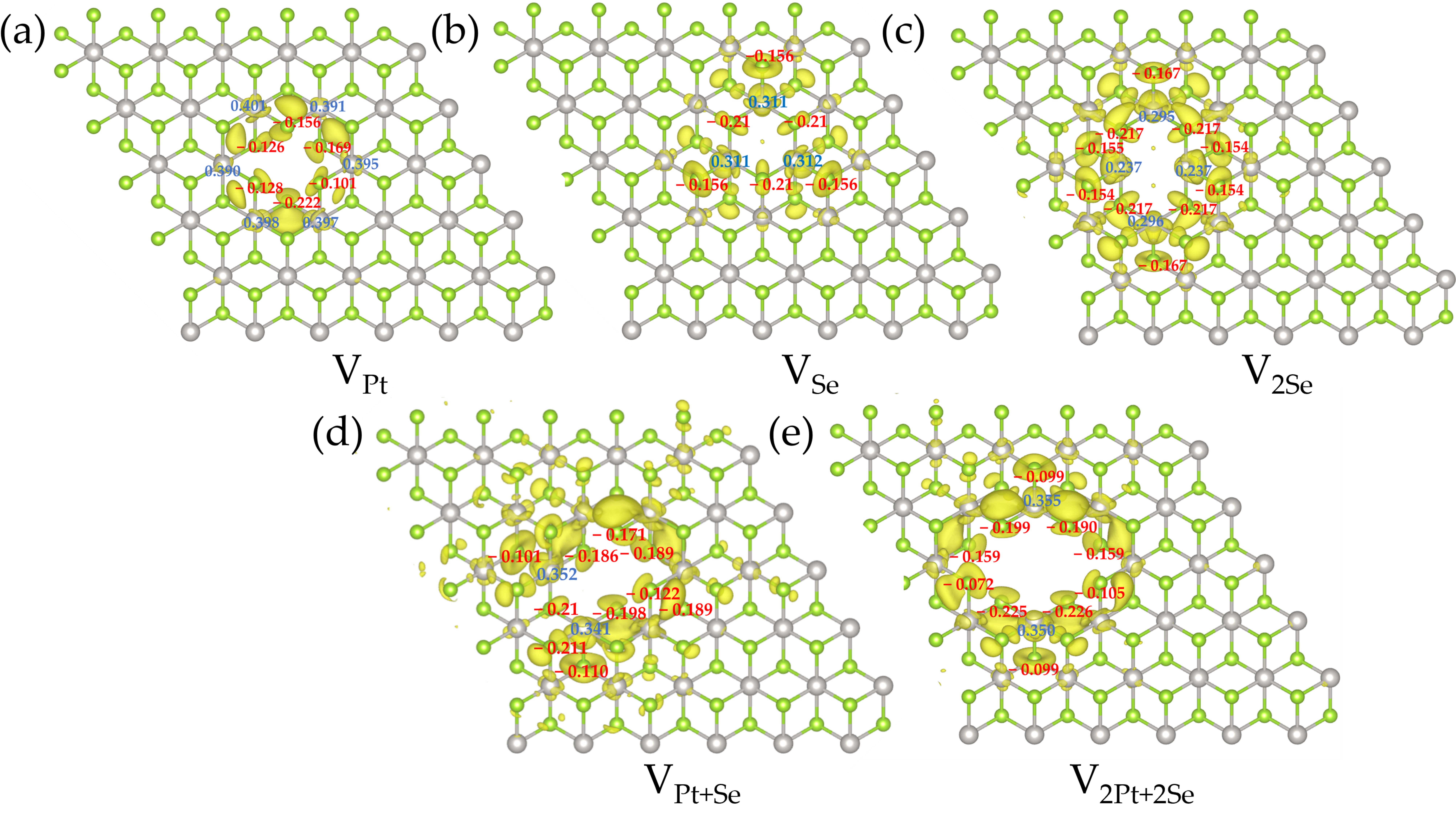}
	\caption{Charge density difference plots of defective PtSe$_2$: 
		(a) V$_{\mathrm{Pt}}$, (b) V$_{\mathrm{Se}}$, (c) V$_{\mathrm{2Se}}$, 
		(d) V$_{\mathrm{Pt}+\mathrm{Se}}$, and (e) V$_{\mathrm{2Pt+2Se}}$. 
		The red numbers denote the charges on Se atoms, whereas the blue numbers denote the charges on Pt atoms.}
	
	\label{fig:fig5}
\end{figure*}

To reveal the mechanisms and physical origins of vacancy-induced electronic-state reconstruction, the TDOS and band structures of monolayer PtSe$_2$ with different defect configurations were systematically analyzed, as shown in Fig.~\ref{fig:fig4} and Fig. S2, respectively. After introducing vacancies, the Fermi level shifts noticeably, and distinct isolated defect peaks appear within the pristine band gap. These in-gap states show very weak dispersion along the high-symmetry paths in the band structures, manifesting as characteristic localized flat bands.

For the cation vacancy defect V$_{\mathrm{Pt}}$, the TDOS shows characteristic acceptor-like states near the VBM (0.1--0.2 eV) and at deeper positions (0.4--0.6 eV); in the corresponding band structure, nearly dispersionless flat bands appear at similar energy positions. For the anion vacancy defects V$_{\mathrm{Se}}$ and V$_{2\mathrm{Se}}$, distinct shallow donor-like characteristics are observed, with sharp defect-state peaks introduced just below the CBM in the 0.9--1.1 eV region. In contrast, the composite defect V$_{\mathrm{Pt+Se}}$ induces a more complex perturbation to the electronic structure; the TDOS shows that the defect levels are widely distributed in both the upper and lower parts of the band gap. Finally, the cluster defect V$_{\mathrm{2Pt+2Se}}$ strongly perturbs the electronic structure of the system, leading to a series of dense deep-level states near the middle of the band gap (0.4--0.8 eV).

The PDOS of all defective systems was further examined to clarify the orbital origins of these defect states (Fig. S3). Comparative analysis shows that the orbital reconstruction of defect states in monolayer PtSe$_2$ follows a consistent trend: the orbital contributions strongly depend on the energy positions of the defect states within the band gap. Defect states near the valence band are dominated by Se-$p$ orbitals, whereas defect states near the conduction band mainly arise from both Pt-$d$ and Se-$p$ orbitals. In particular, for the cation vacancy defect V$_{\mathrm{Pt}}$, the in-gap defect states are mainly contributed by the Se-$p$ orbitals of the neighboring Se atoms, indicating that the Pt vacancy primarily perturbs the local electronic environment of adjacent Se atoms.

The real-space distribution of defect-induced in-gap states was further examined by calculating representative band-decomposed charge densities for selected in-gap bands of different vacancy structures, as shown in Fig. S4. These defect states are mainly distributed around the vacancy regions and their neighboring atoms, confirming their vacancy-induced localized-state character. For the single vacancy defects V$_{\mathrm{Pt}}$ and V$_{\mathrm{Se}}$, the charge density is mainly confined around the atoms adjacent to the vacancy sites. In contrast, for the composite vacancies, including V$_{2\mathrm{Se}}$, V$_{\mathrm{Pt+Se}}$, and V$_{\mathrm{2Pt+2Se}}$, the charge-density distribution becomes more extended, indicating stronger local electronic reconstruction induced by composite vacancy defects.

Vacancy defects in monolayer PtSe$_2$ introduce various deep and shallow flat bands within the band gap. These flat defect bands can act as intermediate states for electron--hole recombination or optical transitions, thereby introducing additional low-energy transition channels and providing a theoretical basis for modulating the optoelectronic response of this material via defect engineering.

Additionally, the magnetic analysis shows that the V$_{\mathrm{Pt}}$ and V$_{\mathrm{2Pt+2Se}}$ defective systems both exhibit pronounced spin-polarized features, with total magnetic moments of approximately $4~\mu_{\mathrm{B}}$. As an auxiliary analysis of spin polarization, spin-polarized calculations without SOC were also performed. The spin-up and spin-down DOS of these two systems are clearly asymmetric, as shown in Fig. S5. In contrast, the spin-up and spin-down channels of the other defective systems remain nearly symmetric, indicating their nonmagnetic nature. Further analysis based on the PDOS and spin-density isosurface plots (Fig. S6) shows that the magnetic moments are mainly distributed on the Se atoms near the vacancies and primarily originate from the asymmetric occupation of Se-$p$ orbitals, while the contribution from neighboring Pt-$d$ orbitals is relatively small.

To further understand the influence of vacancy defects on the local electronic environment, the charge density distribution and CM5 charge analysis were combined to examine the charge redistribution characteristics of different defective systems, as shown in Fig.~\ref{fig:fig5}. Here, the CM5 charges were used as relative descriptors to compare the trends of local charge redistribution around different defects.

In the monolayer PtSe$_2$, the CM5 charges of Pt and Se are approximately $+0.395~e$ and $-0.197~e$, respectively. 
After introducing vacancy defects, the charge distribution of neighboring atoms undergoes significant reconstruction. 
In the V$_{\mathrm{Pt}}$ system, the nearby Se atoms exhibit pronounced polarization: the negative charge of the lower Se increases to $-0.222~e$, indicating electron trapping, while the charges of other Se atoms decrease, suggesting that charge redistribution is mainly localized around adjacent Se atoms, with Pt atoms remaining nearly unchanged. 
This implies that the breaking of Pt--Se coordination bonds primarily induces charge redistribution of Se atoms. 
In the V$_{\mathrm{Se}}$ system, charge accumulation occurs in the neighboring Pt and Se atoms (Pt $\approx +0.31~e$, Se $\approx -0.21~e$), whereas the trans-Se atom opposite to the vacancy along the Pt--Se bonding direction decreases to $-0.156~e$, indicating disruption of $d$--$p$ hybridization and the formation of defect states in the local region. 
For V$_{\mathrm{Pt+Se}}$, the CM5 charge of adjacent Pt atoms decreases from $+0.395~e$ to $+0.34~e$, accompanied by an inhomogeneous redistribution of neighboring Se atoms (from $-0.21~e$ to $-0.12~e$), with the trans-Se further reduced to $-0.10~e$. 
This reveals distinct charge asymmetry and electron localization. 
In the V$_{2\mathrm{Se}}$ system, the CM5 charge of neighboring Pt atoms decreases significantly to $0.237$--$0.296~e$, indicating that the $d$--$p$ hybridization between Pt and Se is weakened. 
Meanwhile, some adjacent Se atoms gain more negative charge ($-0.217~e$), while others bonded to Pt lose charge, reflecting a pronounced polarization effect. 
For the complex vacancy V$_{\mathrm{2Pt+2Se}}$, the CM5 charge of neighboring Pt atoms decreases from $+0.395~e$ to $0.35$--$0.355~e$, suggesting electron accumulation at Pt sites. 
In contrast, the surrounding Se atoms exhibit strongly differentiated charges ranging from $-0.226~e$ to $-0.072~e$, further demonstrating enhanced local polarization and reconstructed $d$--$p$ hybridization. 
These effects introduce strong in-gap defect states and lead to a significant band gap narrowing.

In the previous section, we systematically analyzed the formation energies and electronic structures of neutral vacancies. The results showed that vacancies introduce localized defect states within the band gap and significantly affect carrier-related electronic states and band gap modulation. However, in practical 2D materials, defects are not limited to their neutral form. Charged vacancies are also commonly present and have been experimentally verified. For example, scanning tunneling microscopy (STM) not only induces defect migration but also enables the reversible charging and discharging process of neutral vacancies. Therefore, to further reveal the true stability and physical mechanisms of defects, we extend the study to include various charge states ($+1$, $+2$, $-1$, $-2$) of V$_{\mathrm{Pt}}$ and V$_{\mathrm{Se}}$ in monolayer PtSe$_2$, and calculate their formation energies and stabilities under different chemical potential conditions.

The formation energy of a charged defect is given by
\begin{equation}
	\begin{aligned}
		E_f(q) = & \ E_\mathrm{defect}^{(q)} - E_\mathrm{pristine}^{(0)}
		+ \sum_i n_i \mu_i  \\
		& + q \left(E_\mathrm{VBM} + E_F + \Delta V\right)
		+ E_\mathrm{corr}.
	\end{aligned}
	\label{eq:Ef_charged}
\end{equation}

Here, $E_{\mathrm{defect}}^{(q)}$ and $E_{\mathrm{pristine}}^{(0)}$ represent the total energies of the defect supercell with charge state $q$ and the neutral pristine supercell, respectively; $\Delta V$ denotes the potential alignment term; and $E_{\mathrm{corr}}$ is the finite-size charge correction term. In this work, the finite-size scaling method is employed to treat the charged defect formation energies. Specifically, the formation energies in a series of supercells with different in-plane sizes are systematically calculated and extrapolated to the dilute limit based on their variation with the supercell size. In this finite-size scaling treatment, the electrostatic finite-size interaction and potential-alignment contributions are included in the size-dependent terms and are removed by extrapolation to $L \rightarrow \infty$, where $L$ represents the characteristic in-plane size of the supercell~\cite{54}. Therefore, no separate post-extrapolation $q\Delta V$ or $E_{\mathrm{corr}}$ correction was applied to the final dilute-limit formation energies.
\begin{figure}[b]  
	\centering
	\includegraphics[width=0.45\textwidth]{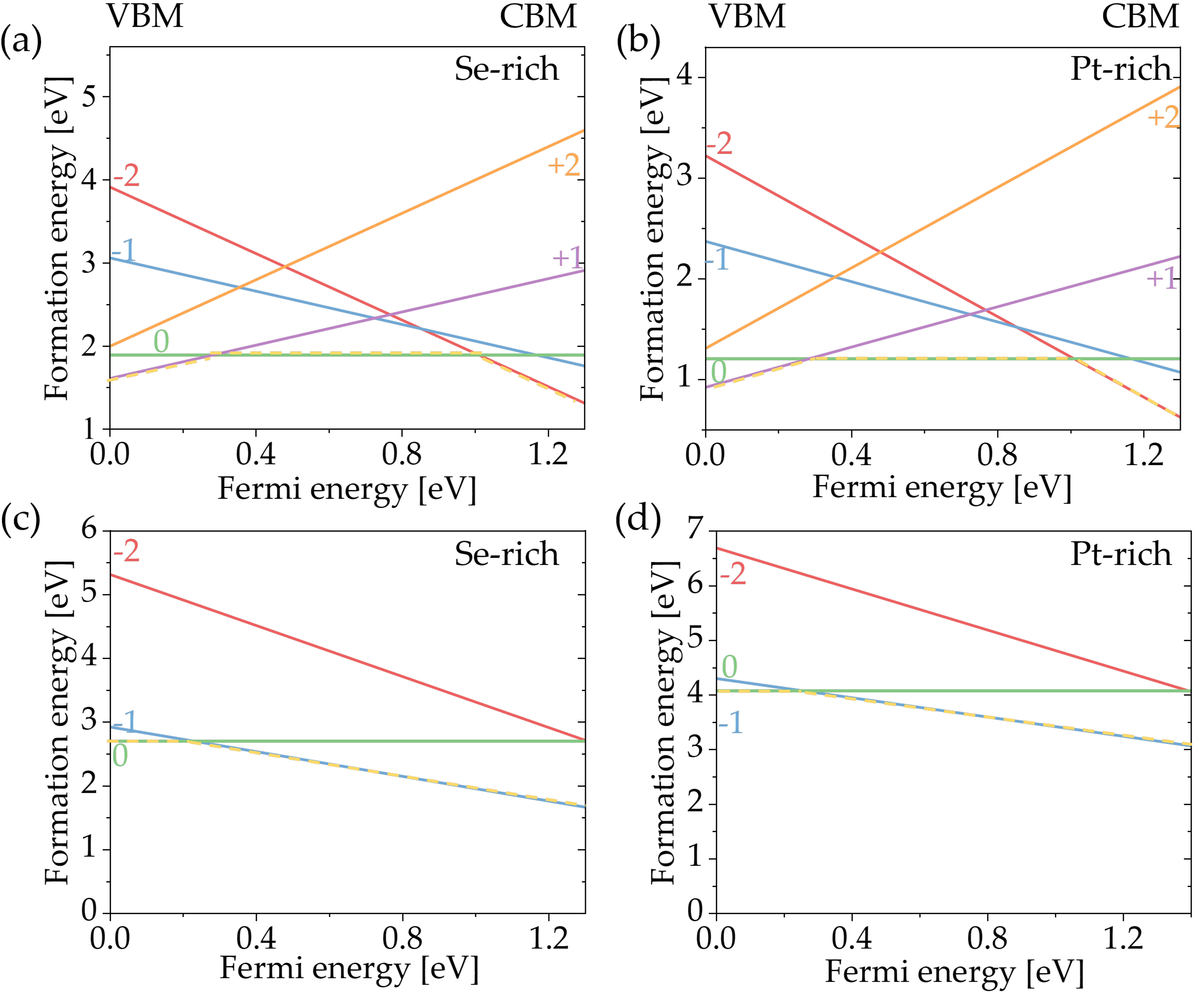}
	\caption{Formation energies of V$_{\mathrm{Se}}$ (a, b) and V$_{\mathrm{Pt}}$ (c, d) defects in monolayer PtSe$_2$ as a function of the Fermi energy under Pt-rich and Se-rich conditions.}
	
	\label{fig:fig6}
\end{figure}

The defect formation energies were obtained by extrapolating the calculated values from different supercell sizes to the infinite-size limit, as shown in Fig.~S7. To further quantify the magnitude of the finite-size effect, the deviations between the \(6\times6\times1\) supercell formation energies and the extrapolated dilute-limit values are summarized in Table~S2 of the Supporting Information. Fig.~\ref{fig:fig6} illustrates the formation energies of intrinsic point defects in monolayer PtSe$_2$ as a function of the Fermi energy under both Se-rich and Pt-rich limit conditions. The yellow dashed lines denote the lowest formation energy envelopes, which represent the thermodynamically most stable charge states at a given Fermi energy and chemical potential. The kinks in these envelopes correspond to the thermodynamic transition levels between different charge states. The thermodynamic charge transition levels $\epsilon(q/q')$ of V$_{\mathrm{Se}}$ and V$_{\mathrm{Pt}}$ are summarized in TABLE~\ref{TabA2}. Furthermore, the formation energies of these defects at the VBM and CBM are listed in TABLE~\ref{TabA3}.

Considering the pronounced amphoteric behavior of V$_{\mathrm{Se}}$, charge states from $-2$ to $+2$ were systematically examined. In contrast, for V$_{\mathrm{Pt}}$, due to the structural instability of high positive charge states, only the range from $-2$ to $0$ is considered. The results indicate that the variation in chemical potential leads to a rigid shift of the formation energy curves. V$_{\mathrm{Se}}$ shows obvious amphoteric characteristics within the band gap. When the Fermi level is near the VBM, the lowest formation energy is mainly dominated by the +1 state. As the Fermi level increases, the neutral state rapidly becomes the thermodynamically most stable state and dominates over most of the band gap region. When the Fermi level further approaches the CBM, the stable charge state changes from the neutral state to the -2 state, indicating that V$_{\mathrm{Se}}$ tends to capture electrons and form a negatively charged state under electron-rich conditions.

In contrast, the charge state evolution of V$_{\mathrm{Pt}}$ exhibits relatively simple acceptor-like characteristics, and it has a lower formation energy under Se-rich conditions. Near the VBM, the lowest formation energy of V$_{\mathrm{Pt}}$ is dominated by the neutral state. As the Fermi level slightly increases, the -1 state rapidly becomes the most stable charge state and remains dominant over most of the band gap region up to the conduction band minimum. Therefore, V$_{\mathrm{Pt}}$ mainly behaves as an acceptor-like vacancy defect, with stable charge states mainly distributed between 0 and -1.
Therefore, by synergistically tuning the chemical potential during growth and the Fermi energy, the conductivity type and potential Fermi level pinning effects in PtSe$_2$ can be effectively modulated.
\begin{table}[tb]
	\caption{\label{TabA2}Thermodynamic charge transition levels $\epsilon(q/q')$ of charged vacancies in a PtSe$_2$ monolayer. The values are referenced to the VBM.}
	\renewcommand{\arraystretch}{1.5}
	\begin{ruledtabular}
		\begin{tabular}{lcc}
			Defect & Transition level & $\epsilon(q/q')$ (eV) \\
			\colrule
			V$_{\mathrm{Se}}$ & $\epsilon(+1/0)$ & 0.283 \\
			V$_{\mathrm{Se}}$ & $\epsilon(0/-2)$ & 1.010 \\
			V$_{\mathrm{Pt}}$ & $\epsilon(0/-1)$ & 0.223 \\
		\end{tabular}
	\end{ruledtabular}
\end{table}
\begin{table}[tb]
	\caption{\label{TabA3}Formation energies of charged vacancies in the PtSe$_2$ monolayer (eV).}
	\renewcommand{\arraystretch}{1.5}
	\begin{ruledtabular}
		\begin{tabular*}{\linewidth}{@{\extracolsep{\fill}}l l cccc cc}
			Type of vacancy &        & \multicolumn{4}{c}{ V$_{\mathrm{Se}}$}        & \multicolumn{2}{c}{ V$_{\mathrm{Pt}}$} \\
			\cline{3-6}\cline{7-8}
			Charge, $e$     &        & +2 & +1 & $-1$ & $-2$         & $-1$                    & $-2$ \\
			\colrule
			\multirow{2}{*}{Se-rich}  & VBM & 1.99 & 1.61 & 3.06 & 3.91 & 2.92  & 5.31 \\
			& CBM & 4.63 & 2.93 & 1.74 & 1.28 & 1.65 & 2.68   \\
			\multirow{2}{*}{Pt-rich}  & VBM & 1.31 & 0.92 & 2.37 & 3.22 & 4.30 & 6.69   \\
			& CBM & 3.95 & 2.24 & 1.05 & 0.59 & 3.07 & 4.06 \\
		\end{tabular*}
	\end{ruledtabular}
\end{table}

\begin{figure}[tb] 
	\centering
	\includegraphics[width=0.45\textwidth]{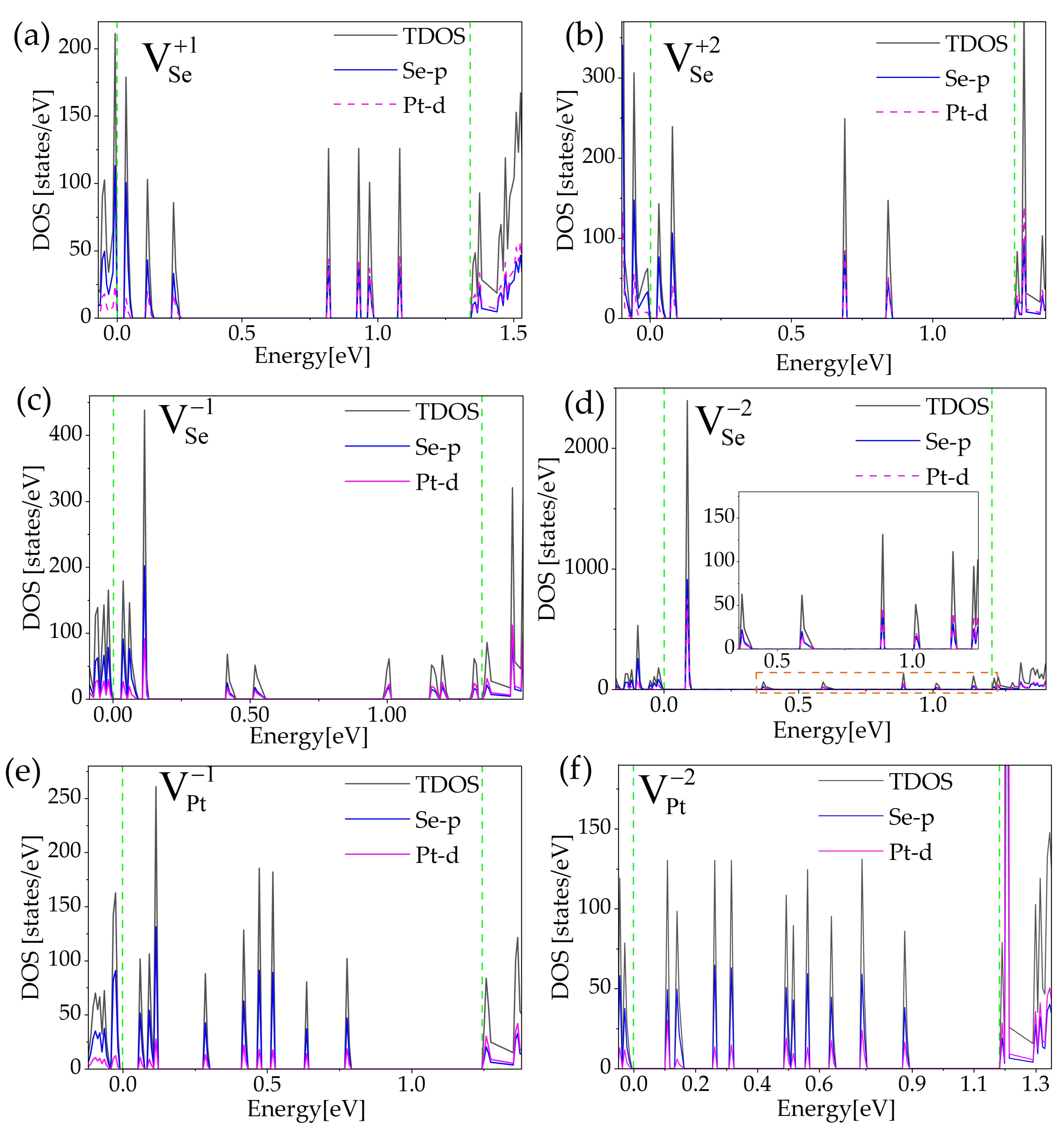}
	\caption{PDOS of charged vacancy defects in monolayer PtSe$_2$ calculated with SOC: 
		(a) \(V_{\mathrm{Se}}^{+2}\), (b) \(V_{\mathrm{Se}}^{+1}\), 
		(c) \(V_{\mathrm{Se}}^{-1}\), (d) \(V_{\mathrm{Se}}^{-2}\), 
		(e) \(V_{\mathrm{Pt}}^{-1}\), and (f) \(V_{\mathrm{Pt}}^{-2}\). 
		The Fermi level is set to 0 eV.}
	
	\label{fig:fig7}
\end{figure}

The orbital contributions of the charged defect states were evaluated using the PDOS, as shown in Fig.~\ref{fig:fig7} for V$_{\mathrm{Se}}$ and V$_{\mathrm{Pt}}$ in different charge states. The results show that the orbital-contribution characteristics of the charged systems are generally consistent with those of the neutral defect systems, namely, the orbital origins of the defect states remain closely related to their energy positions within the band gap. For the charged V$_{\mathrm{Se}}$ systems, the defect peaks near the VBM mainly originate from Se-$p$ orbitals, whereas the defect states near the CBM are jointly contributed by Pt-$d$ and Se-$p$ orbitals, with relatively comparable contributions from both components (Fig.~\ref{fig:fig7}(a-d)). For different charge states of V$_{\mathrm{Pt}}$, the in-gap defect levels are mainly contributed by Se-$p$ orbitals near the vacancy, indicating that Pt vacancies mainly regulate the in-gap states by reconstructing the local electronic environment of the neighboring Se atoms (Fig.~\ref{fig:fig7}(e-f)).

Further analysis of the TDOS shows that, except for V$_{\mathrm{Se}}^{+2}$, most charged defect systems introduce more defect peaks within the band gap compared with the neutral state. Accompanied by the reconstruction of these defect levels, the magnetism of the systems also changes significantly: except for V$_{\mathrm{Se}}^{+2}$, which is nonmagnetic, the other charged defect systems exhibit net magnetic moments to different degrees.

The physical origin of the magnetic moments was further clarified using spin-polarized DOS calculations without SOC as an auxiliary analysis for each system (Fig. S8). For the V$_{\mathrm{Se}}$ systems, the energy-level distribution and degree of localization show a clear dependence on the charge state. Near the CBM, V$_{\mathrm{Se}}^{-2}$ exhibits pronounced spin-polarized features: a distinct impurity peak appears in the spin-up channel, whereas the spin-down channel is strongly suppressed, resulting in a magnetic moment of approximately $2~\mu_{\mathrm{B}}$. For the V$_{\mathrm{Se}}^{-1}$ state, although the intensity of the in-gap states is relatively weak, spin polarization is still retained, inducing a net magnetic moment of approximately $1~\mu_{\mathrm{B}}$. In contrast, for positively charged V$_{\mathrm{Se}}$, V$_{\mathrm{Se}}^{+1}$ possesses a magnetic moment of approximately $0.9~\mu_{\mathrm{B}}$, whereas the defect states in V$_{\mathrm{Se}}^{+2}$ tend to be symmetrically occupied, leading to spin cancellation and rendering the system nonmagnetic.

For negatively charged V$_{\mathrm{Pt}}$, the shallow acceptor states are mainly concentrated near the VBM, while the dominant defect states are located deep within the band gap. Charge induced spin splitting gives V$_{\mathrm{Pt}}^{-1}$ a magnetic moment of approximately $3~\mu_{\mathrm{B}}$. Even in the $-2$ charge state, the system still maintains a stable spin-polarized configuration and retains a magnetic moment of approximately $2~\mu_{\mathrm{B}}$. Therefore, charged vacancies not only modify the distribution of defect levels, but also induce pronounced spin-polarized behavior by regulating the occupation of defect states.

\begin{figure}[b] 
	\centering
	\includegraphics[width=0.45\textwidth]{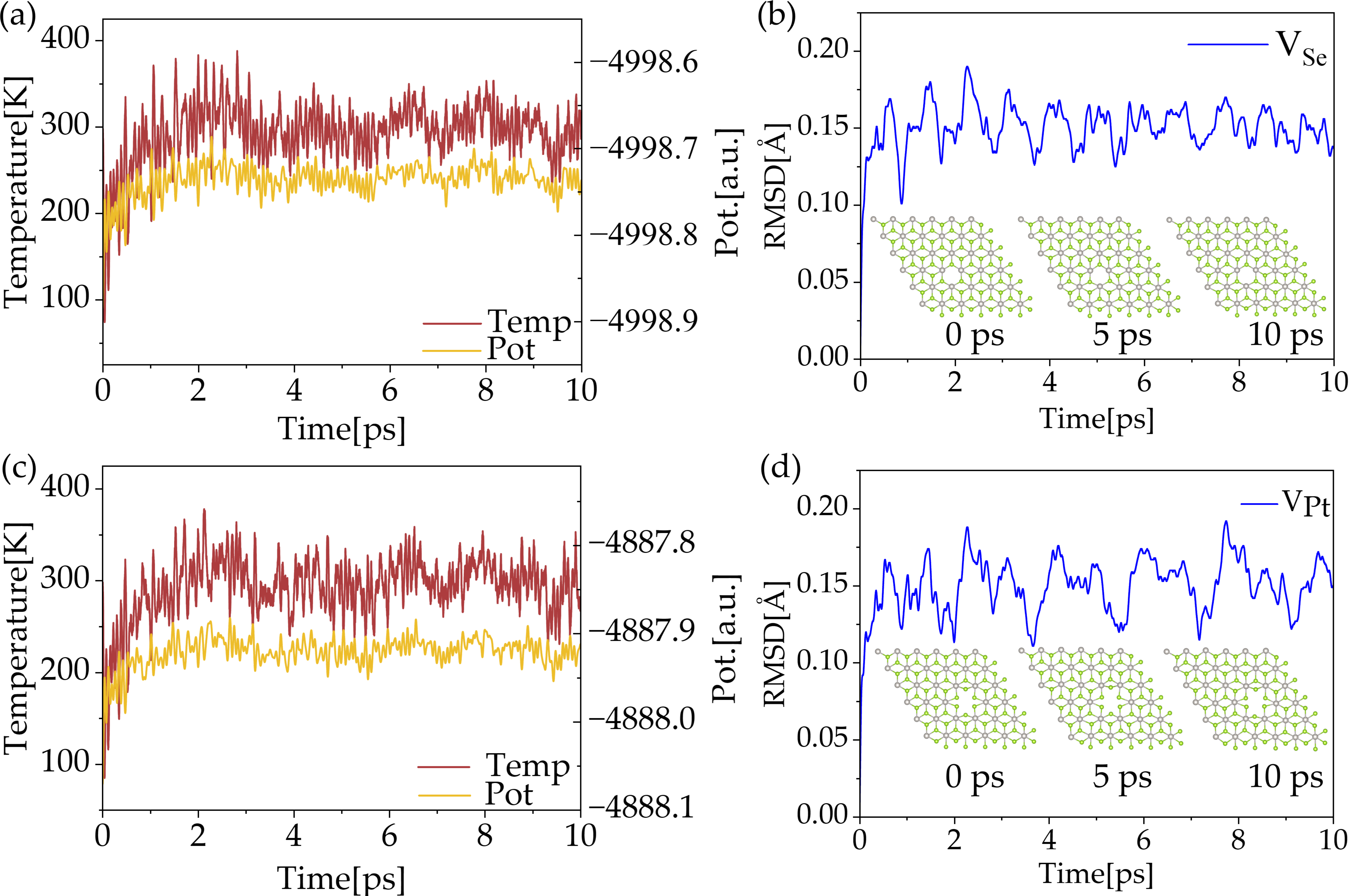}
	\caption{AIMD simulations of V${_\mathrm{Se}}$ and V${_\mathrm{Pt}}$ in the PtSe$_2$ monolayer at 300 K. (a, c) Energy and temperature evolution curves of V${_\mathrm{Se}}$ and V${_\mathrm{Pt}}$ with respect to time. (b, d) RMSD evolution over time and structural snapshots at different time points.}
	\label{fig:fig8}
\end{figure}

To provide a short-time finite-temperature check of the structural robustness of defects in monolayer PtSe$_2$, AIMD simulations were performed for V$_{\mathrm{Pt}}$ and V$_{\mathrm{Se}}$ at 300 K. It should be noted that the accessible AIMD time scale is limited; therefore, these simulations are not intended to establish long-term thermodynamic stability, but rather to examine whether the defective structures undergo immediate structural collapse, defect healing, or obvious vacancy migration within the simulated time window.
As shown in Fig.~\ref{fig:fig8}, at 300 K, both the energy and temperature evolution curves obtained from the AIMD simulations of V$_{\mathrm{Pt}}$ and V$_{\mathrm{Se}}$ exhibit stable fluctuations after equilibration. After an initial rapid fluctuation, the system temperature quickly stabilizes near the target value and remains within a reasonable thermal fluctuation range. The total and potential energies also oscillate around their respective average values without noticeable drift, indicating that the system maintains stable thermal fluctuations within the AIMD time window.
Further comparison of structural snapshots at different time segments reveals that the local environment around the defects remains essentially unchanged, showing no significant reconstruction or healing behavior. For both types of vacancies, although neighboring atoms undergo slight structural relaxation induced by the defects, the vacancy positions remain unchanged throughout the simulation, and no signs of migration are observed. This indicates that both vacancies do not undergo immediate structural collapse, defect healing, or obvious migration within the simulated time window at 300 K. 
Root Mean Square Deviation (RMSD) analysis further supports the above conclusions. The RMSD curves of both defective systems exhibit a characteristic rapid rise followed by a plateau-like oscillation, supporting the short-time dynamical stability of the defective structures at room temperature. A quantitative comparison shows that V$_{\mathrm{Se}}$ exhibits a higher RMSD plateau value and larger fluctuation amplitude compared with V$_{\mathrm{Pt}}$, reflecting that the bonding environment around Se vacancies is more flexible and prone to thermally induced vibrations, whereas the region surrounding Pt vacancies is more rigid, and the local structure is less susceptible to perturbation.

To ensure the identification of a reasonable minimum-barrier mechanism, the possible migration networks of the vacancy defects were evaluated, and the minimum energy path (MEP) was obtained by performing PBE+SOC single-point energy calculations on the converged NEB images. The CI-NEB calculations used 5 intermediate images for the V$_{\mathrm{Se}}$ migration pathways and 6 intermediate images for the V$_{\mathrm{Pt}}$ migration pathway, excluding the initial and final states.

\begin{figure}[t] 
	\centering
	\includegraphics[width=0.45\textwidth]{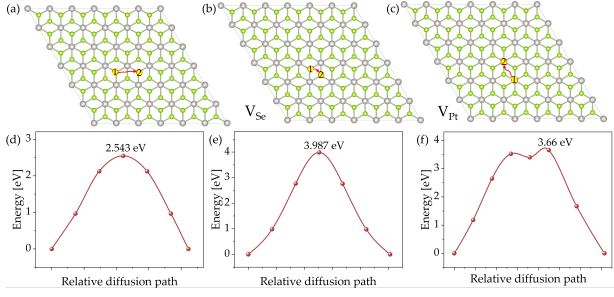}
	\caption{Schematic diagrams of the initial state (1) and final state (2) for the vacancy diffusion pathways in monolayer PtSe$_2$: (a) in-plane migration of V$_{\mathrm{Se}}$, (b) proximal inter-layer migration of V$_{\mathrm{Se}}$, and (c) migration of V$_{\mathrm{Pt}}$. (d-f) The corresponding energy barrier profiles for the vacancy diffusion pathways shown in (a-c).}
	\label{fig:fig9}
\end{figure}

Fig.~\ref{fig:fig9}(a, d) and (b, e) show the in-plane migration and proximal inter-layer migration processes of V$_{\mathrm{Se}}$, respectively. Calculations indicate that the energies of the initial and final states are nearly identical, suggesting that the defect sites are thermodynamically equivalent. Among them, the barrier for the in-plane migration mechanism is relatively lower, at approximately 2.543 eV (Fig.~\ref{fig:fig9}(d)). In contrast, the proximal inter-layer migration process (approximately 3.987 eV, Fig.~\ref{fig:fig9}(e)) involves significant structural reconstruction, during which the surrounding Pt--Se bonds undergo intense distortion and rearrangement with a high bonding cost. This indicates that the diffusion of V$_{\mathrm{Se}}$ is primarily restricted to in-plane migration. 

Fig.~\ref{fig:fig9}(c, f) displays the migration results for V$_{\mathrm{Pt}}$, with a main saddle point at approximately 3.66 eV. Notably, the MEP exhibits a plateau region in the range of 3.4--3.52 eV before crossing the main saddle point, implying that the diffusion process of V$_{\mathrm{Pt}}$ involves a complex, synergistic mechanism of multi-bond breaking and reconstruction. 

Although the in-plane migration path with a lower barrier was explicitly identified, the minimum migration barriers for both V$_{\mathrm{Se}}$ (2.543 eV) and V$_{\mathrm{Pt}}$ (3.66 eV) remain extremely high, far exceeding the thermal fluctuation energy at room temperature. These calculated high migration barriers and complex transition mechanisms are consistent with the theoretical results previously reported by Gao et al. regarding the diffusion behavior of intrinsic vacancies in PtSe$_2$~\cite{29}. Therefore, it is fundamentally difficult for these two types of vacancies to undergo spontaneous migration at normal operating temperatures, exhibiting high kinetic stability.

To further reveal the influence of point defects on the optical response of monolayer PtSe$_2$, its frequency-dependent linear optical properties were calculated and analyzed. In two-dimensional models, the size of the vacuum layer in the supercell significantly interferes with the accurate evaluation of the dielectric function $\epsilon(\omega)$ and related optical parameters (such as optical absorption and refraction behavior). To avoid the ambiguity caused by the artificial vacuum thickness, a more fundamental, purely 2D physical descriptor, namely the electronic polarizability $\alpha(\omega)$, is adopted in this work. Tian et al.~\cite{55} confirmed that the real part $\alpha_1(\omega)$ and the imaginary part $\alpha_2(\omega)$ of the polarizability mainly depend on the intrinsic electronic transitions of the material, which are independent of the vacuum layer thickness after proper conversion. Based on this advantage, the optical response analysis for all systems in this study is performed by converting to the electronic polarizability, thereby enabling a more reliable comparison among different defective systems.

\begin{figure}[t]  
	\centering
	\includegraphics[width=0.45\textwidth]{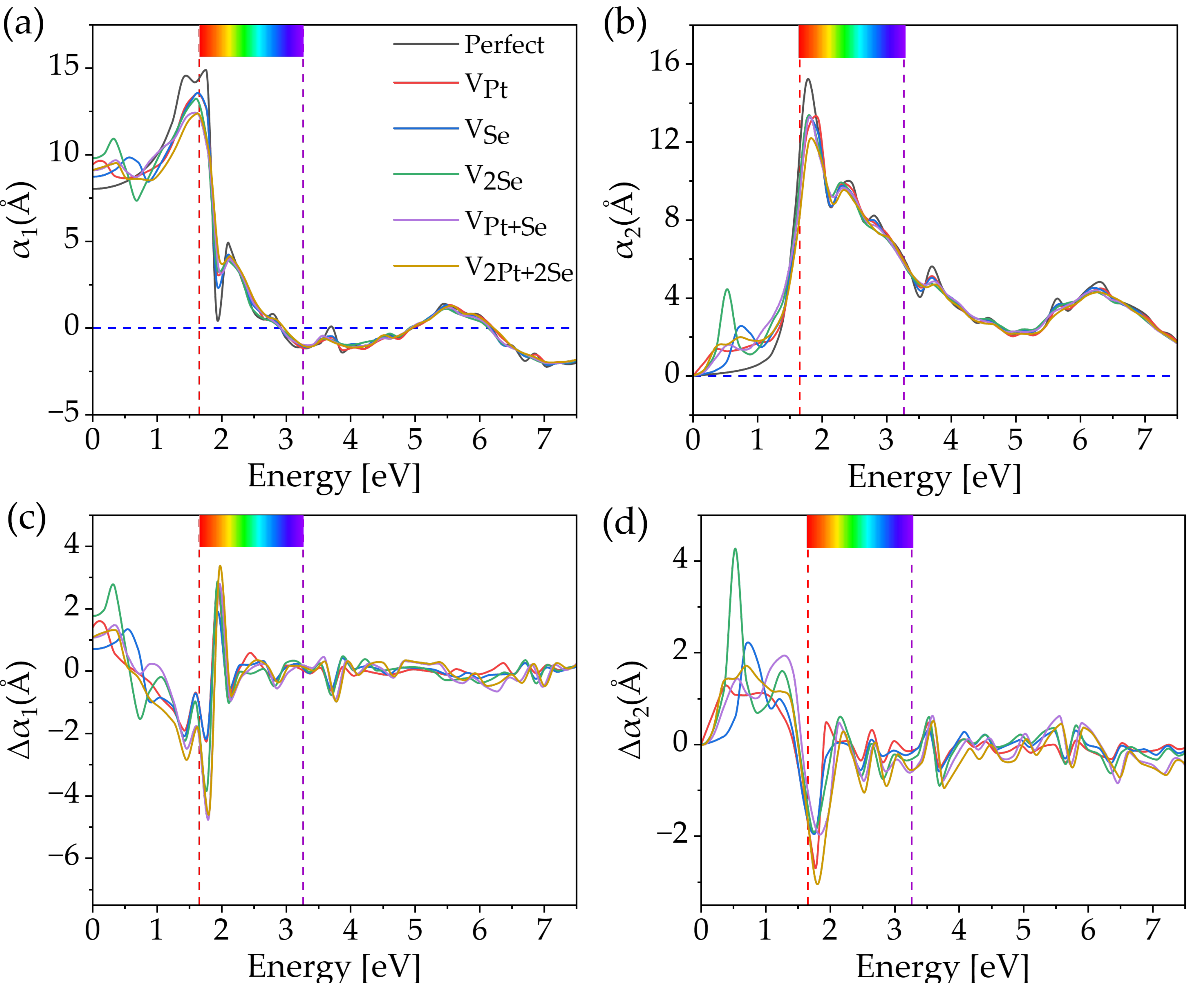}
	\caption{Optical polarizability along the $x$-direction of pristine monolayer $\text{PtSe}_2$ and its five point-defect structures as a function of photon energy. (a) Real part $\alpha_1$ and (b) imaginary part $\alpha_2$ of the polarizability. (c) Differential real part $\Delta\alpha_1$ and (d) differential imaginary part $\Delta\alpha_2$ of the polarizability induced by the defects. The colored bar at the top indicates the visible spectral range.}
	\label{fig:fig10}
\end{figure}

Fig.~\ref{fig:fig10} displays the real part $\alpha_1$ and imaginary part $\alpha_2$ of the polarizability along the $x$-direction for each system, as well as the additional differential responses $\Delta\alpha_1$, $\Delta\alpha_2$ introduced by the defects. The real part of the polarizability, $\alpha_1(\omega)$, reflects the optical dispersion and dielectric screening capability of the material for in-plane electric-field polarization. As shown in Fig.~\ref{fig:fig10}(a), pristine PtSe$_2$ exhibits the highest $\alpha_1$ peak near the main visible-light excitation region at approximately 1.8 eV, whereas the composite vacancy systems show relatively lower peak intensities. After this peak, $\alpha_1$ decreases rapidly for all systems. Among them, the pristine system shows the steepest decrease, while the $V_{\mathrm{2Pt+2Se}}$ system shows the slowest decrease. Subsequently, $\alpha_1$ rises again in the nearby energy region and forms a smaller peak for all systems, with the pristine system still exhibiting the highest peak intensity. The differential spectra in Fig.~\ref{fig:fig10}(c) further show that all defective systems exhibit pronounced negative valleys around 1.8 eV, indicating that vacancy defects weaken the intrinsic dielectric response of pristine PtSe$_2$ associated with the main visible-light transition. This suggests that the defects break the periodic symmetry of the lattice and thus have a significant influence on the in-plane dielectric response. Among them, the negative features of $V_{\mathrm{Pt+Se}}$ and $V_{\mathrm{2Pt+2Se}}$ are the most prominent, indicating that composite vacancies induce stronger perturbations to the in-plane dielectric response. In the energy range of approximately 2.9--4.9 eV, $\alpha_1$ gradually decreases to negative values. In the broader mid-to-high energy region, the $\alpha_1$ curves and fluctuation amplitudes of different systems show only relatively small differences, indicating that the dielectric dispersion in this energy region is mainly determined by the overall band framework and is less sensitive to local vacancy defects.

The imaginary part of the polarizability, $\alpha_2(\omega)$, and the differential imaginary part, $\Delta\alpha_2(\omega)$, along the $x$-direction reflect the absorption-related optical response of the material for photons of different energies. As shown in Fig.~\ref{fig:fig10}(b), pristine monolayer PtSe$_2$ exhibits a weak optical response in the low-energy region. After introducing vacancy defects, the localized states within the band gap provide additional transition channels for low-energy photons, thereby enhancing the low-energy absorption-related response. In particular, $V_{2\mathrm{Se}}$ exhibits a sharp positive response feature around 0.5 eV, indicating a distinct defect-induced infrared absorption response. Near the main absorption peak at approximately 1.8 eV, pristine PtSe$_2$ still shows the highest $\alpha_2$ intensity, whereas $V_{\mathrm{2Pt+2Se}}$ shows the lowest main peak intensity among the considered systems. The $\Delta\alpha_2$ spectra in Fig.~\ref{fig:fig10}(d) further show that all defective systems exhibit positive differential responses in the low-energy range of 0--1.5 eV, but pronounced negative features near 1.8 eV. This indicates that the low-energy absorption channels introduced by defects are accompanied by a weakening of the intrinsic main visible-light absorption transition of pristine PtSe$_2$. As the photon energy further increases, the absorption-related responses of different systems gradually converge, consistent with the mid-to-high energy trend reflected by $\alpha_1$.

In summary, different types of vacancy defects not only significantly alter the intrinsic dielectric response of monolayer PtSe$_2$, but also effectively activate the material's optical absorption in the infrared to near-infrared regions by introducing localized states within the band gap. Furthermore, the optical response analysis along the $y$-direction (detailed in Fig. S9) yields the same overall qualitative conclusions, with only numerical differences existing among the various systems.

As a supplementary application-oriented descriptor, the effect of vacancy-induced electronic reconstruction on local H adsorption thermodynamics was further examined. The Se site adjacent to the V$_{\mathrm{2Pt+2Se}}$ defect exhibits a $\Delta G_{H^*}$ value of 0.025 eV, which is close to the thermoneutral value and closer to 0 eV than the Pt(111) descriptor benchmark of $-0.09$ eV~\cite{56}. The detailed analysis is provided in Fig. S10 of the Supporting Information.

\section{Conclusion}
In summary, by combining defect formation thermodynamics, charge state evolution, electronic structure reconstruction, CI-NEB calculations, short AIMD simulations, and optical response, this work develops an integrated physical framework that captures the behavior of intrinsic vacancy defects in monolayer PtSe$_2$. The calculated results indicate that the formation energies of vacancy defects are highly sensitive to both chemical potential conditions and the Fermi level. Specifically, V$_{\mathrm{Se}}$ is identified as the most favorable vacancy defect throughout the entire chemical potential range, followed by V$_{\mathrm{2Se}}$ as the secondary stable state under Pt-rich conditions. As the growth environment shifts from Pt-rich to Se-rich conditions, the formation energies of Se vacancies, including V$_{\mathrm{Se}}$ and V$_{2\mathrm{Se}}$, increase, whereas those of Pt-related vacancies decrease to different extents, indicating that the chemical potential can effectively regulate the formation tendency of different types of vacancies.
Furthermore, the stability ranges of charge states for each defect, along with the corresponding thermodynamic charge transition levels, are clearly defined by the evolution of formation energies with respect to the Fermi level. Analyses of the density of states and local electronic structure further reveal that significant in-gap defect states are introduced by vacancies; this leads to marked alterations in the orbital hybridization and charge distribution of neighboring atoms, thereby reconstructing the local electronic environment and subsequently influencing the dielectric response. Optical property calculations show that point defects also generate new low-energy absorption channels associated with in-gap defect states. Overall, this study demonstrates that monolayer PtSe$_2$ is a highly tunable 2D platform in which defect engineering enables coordinated modulation of electronic, optical, and adsorption-related properties.
\section{Acknowledgements}
This work was supported by National Natural Science Foundation of China (Grant No. 12474218), Beijing Natural Science Foundation (Grant Nos. 1242022 and 1252022), and Basic Research Project (Innovation Development Project) of Liaoning Provincial Department of Education (Grant No. LJ242510148005).
\section{DATAAVAILABILITY}
The data that support the findings of this article are openly available~\cite{57}.

\bibliography{main}

\clearpage

\newpage
\noindent
\clearpage
\setcounter{figure}{0}\setcounter{table}{0}\setcounter{equation}{0}
\renewcommand{\thefigure}{S\arabic{figure}}
\renewcommand{\thetable}{S\arabic{table}}
\renewcommand{\theequation}{S\arabic{equation}}
\onecolumngrid

\begin{center}{\large\bfseries Supplementary Materials for ``Vacancy-Driven Electronic Reconstruction in Monolayer PtSe$_2$: Formation Thermodynamics and Charge States''}\end{center}

\begin{table}[H]
	\caption{Comparison between previous studies on vacancy defects in PtSe$_2$ and the present work.}
	\label{tab:comparison_previous_work}
	\centering
	\scriptsize
	\renewcommand{\arraystretch}{1.25}
	
	\begin{tabular}{llll}
		\hline
		\parbox[t]{2cm}{\textbf{Reference}} &
		\parbox[t]{4.2cm}{\textbf{Method / SOC treatment}} &
		\parbox[t]{5.3cm}{\textbf{Main work}} &
		\parbox[t]{6.0cm}{\textbf{Key conclusions}} \\
		\hline
		
		\parbox[t]{2cm}{Gao et al.~\citeSI{1}} &
		\parbox[t]{4.2cm}{VASP-PBE; spin-polarized DFT; CI-NEB} &
		\parbox[t]{5.3cm}{Studied neutral Se/Pt single and double vacancies in monolayer PtSe$_2$, including structures, electronic properties, magnetism, and migration barriers.} &
		\parbox[t]{6.0cm}{Se and Pt vacancies significantly modify the electronic structure. Pt single and double vacancies induce pronounced spin polarization. The Pt-vacancy migration barrier is about 2.56~eV, and the lowest Se-vacancy migration barrier is about 2.17~eV.} \\
		\hline
		
		\parbox[t]{2cm}{Zheng et al.~\citeSI{2}} &
		\parbox[t]{4.2cm}{STM/STS; VASP-LDA} &
		\parbox[t]{5.3cm}{Identified intrinsic point defects in ultrathin 1T-PtSe$_2$.} &
		\parbox[t]{6.0cm}{Se vacancies, Pt vacancies, and Se$_{\mathrm{Pt}}$ antisite defects were identified. Se$_{\mathrm{Pt}}$ antisites are relatively abundant under Se-rich conditions.} \\
		\hline
		
		\parbox[t]{2cm}{Shawkat et al.~\citeSI{3}} &
		\parbox[t]{4.2cm}{CVD growth and Ar-plasma treatment; DFT band-structure calculations} &
		\parbox[t]{5.3cm}{Investigated plasma-induced Se defects and their effects on transport in few-layer PtSe$_2$.} &
		\parbox[t]{6.0cm}{Se defects introduce midgap states and reduce the band gap. Ar-plasma treatment converts few-layer PtSe$_2$ from semiconducting to metallic transport.} \\
		\hline
		
		\parbox[t]{2cm}{Avsar et al.~\citeSI{4}} &
		\parbox[t]{4.2cm}{Magnetotransport experiments, STEM, and VASP-PBE-SOC} &
		\parbox[t]{5.3cm}{Studied Pt-vacancy-induced magnetism in monolayer and bilayer PtSe$_2$.} &
		\parbox[t]{6.0cm}{Pt vacancies are the main origin of defect-induced magnetism. Monolayer PtSe$_2$ shows antiferromagnetic behavior, while bilayer PtSe$_2$ shows ferromagnetic behavior.} \\
		\hline
		
		\parbox[t]{2cm}{Chang et al.~\citeSI{5}} &
		\parbox[t]{4.2cm}{First-principles calculations and HER/OER experiments; water solvation considered; SOC not emphasized} &
		\parbox[t]{5.3cm}{Studied the HER/OER catalytic performance of Se-vacancy-engineered PtSe$_2$.} &
		\parbox[t]{6.0cm}{Se vacancies are kinetically stable and significantly enhance HER/OER activity. Se-1V, Se-2V, and Se-3V all show favorable HER activity.} \\
		\hline
		
		\parbox[t]{2cm}{Present work} &
		\parbox[t]{4.2cm}{VASP-PBE; SOC included for electronic structures, defect states, charged-defect formation energies, thermodynamic transition levels, optical properties, H adsorption energetics, and SOC-corrected NEB energy profiles; CP2K-AIMD; CI-NEB} &
		\parbox[t]{5.3cm}{Systematically studied neutral and charged intrinsic vacancies in monolayer PtSe$_2$, including formation thermodynamics, electronic structure, kinetic stability, and optical response.} &
		\parbox[t]{6.0cm}{A Fermi-level-dependent charged-defect thermodynamic framework is established for $V_{\mathrm{Se}}$ and $V_{\mathrm{Pt}}$. Stable charge-state regions are clarified, and charge-state effects, kinetic stability, and optical response are connected within one framework.} \\
		\hline
		
	\end{tabular}
\end{table}
\begin{figure}[H]
	\centering
	\includegraphics[width=0.95\textwidth]{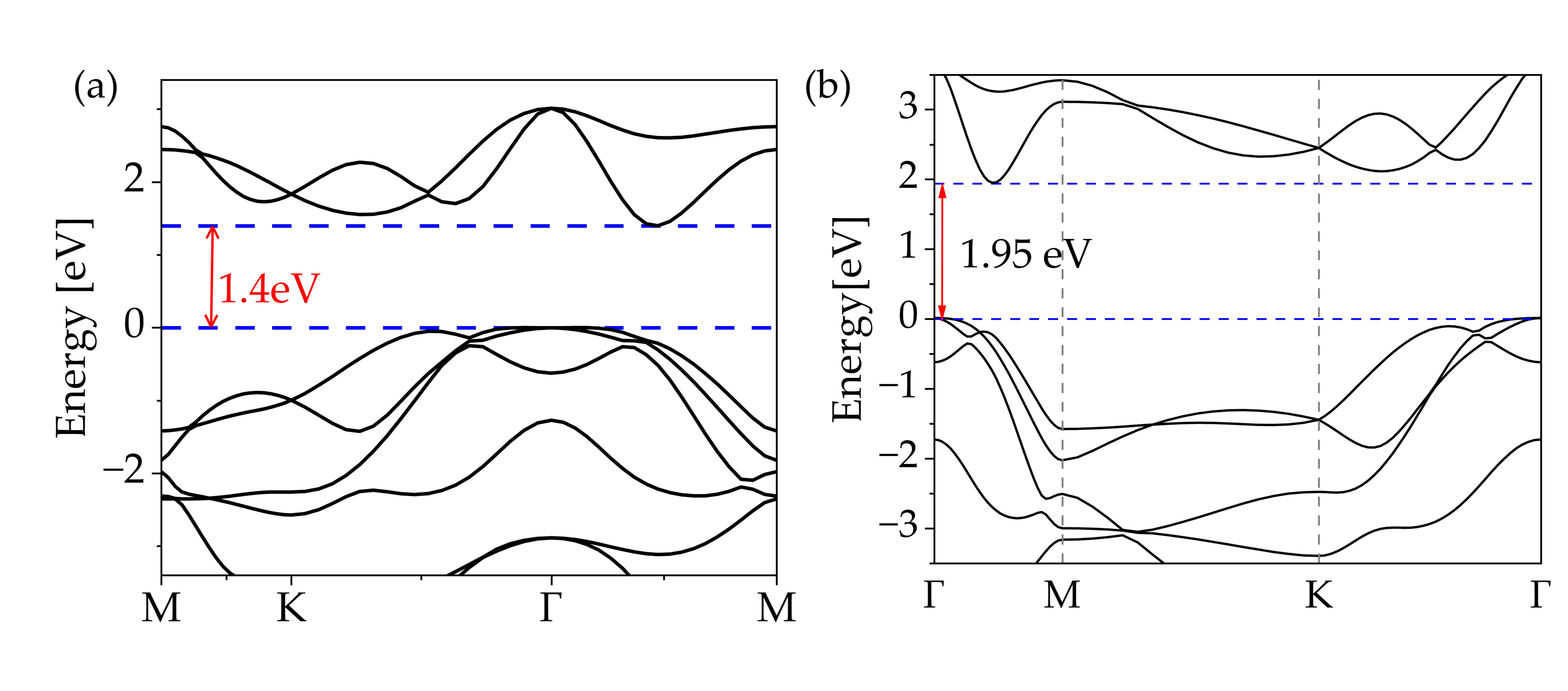}
	\caption{Electronic band structures of pristine monolayer PtSe$_2$ calculated using the (a) PBE and (b) HSE06 methods.}
	
	\label{fig:figS1}
\end{figure}
\begin{figure}[!htbp]
	\centering
	\includegraphics[width=0.95\textwidth]{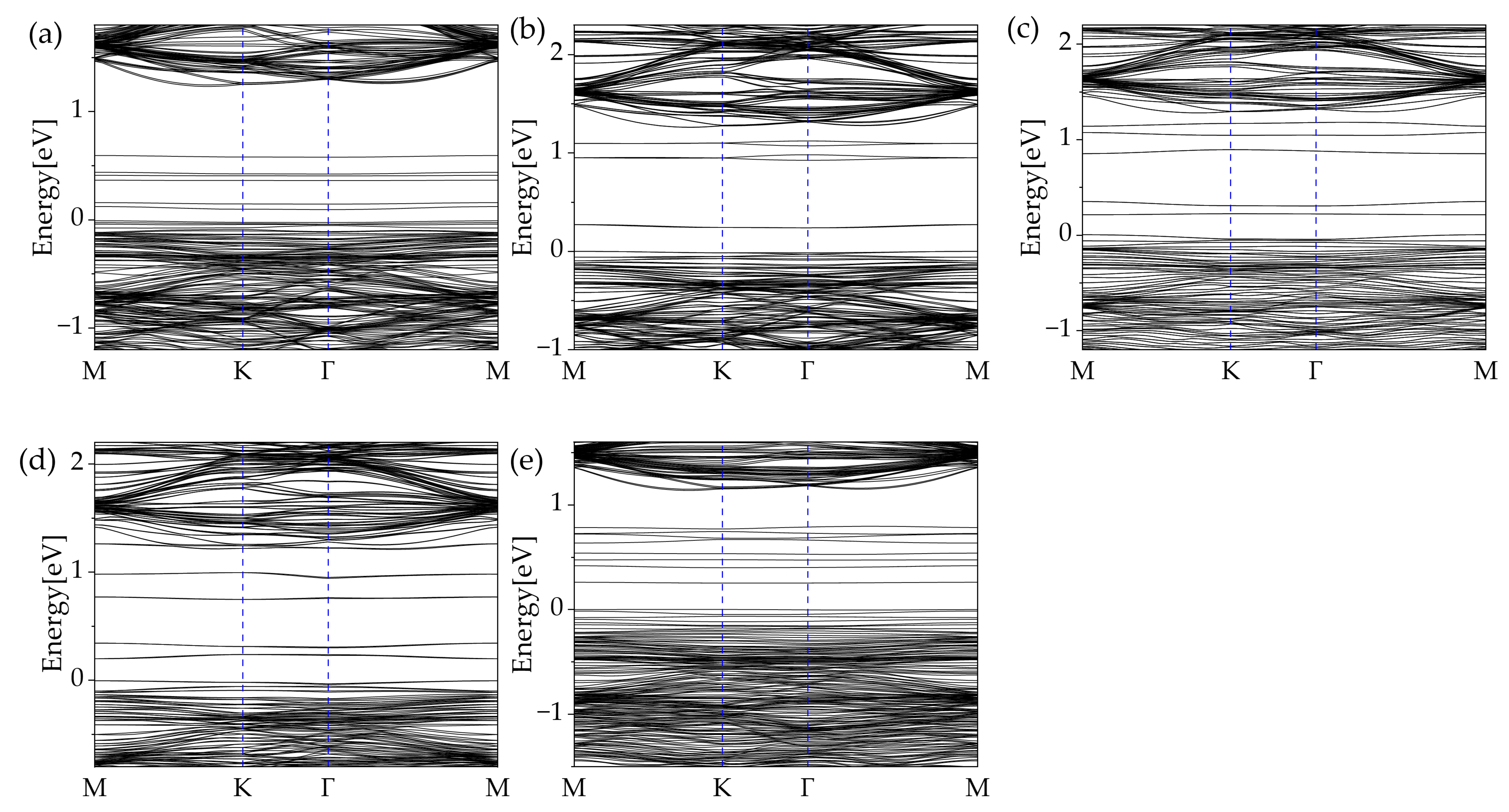}
	\caption{Electronic band structures of defective PtSe$_2$: 
		(a) V$_{\mathrm{Pt}}$, (b) V$_{\mathrm{Se}}$, (c) V$_{\mathrm{2Se}}$, 
		(d) V$_{\mathrm{Pt}+\mathrm{Se}}$, and (e) V$_{2\mathrm{Pt}+2\mathrm{Se}}$.}
	
	\label{fig:figS2}
\end{figure}

\begin{figure}[H]
	\centering
	\includegraphics[width=0.95\textwidth]{figS3}
	\caption{PDOS of defective PtSe$_2$: 
		(a) V$_{\mathrm{Pt}}$, (b) V$_{\mathrm{Se}}$, (c) V$_{\mathrm{2Se}}$, 
		(d) V$_{\mathrm{Pt}+\mathrm{Se}}$, and (e) V$_{2\mathrm{Pt}+2\mathrm{Se}}$.}
	
	\label{fig:figS3}
\end{figure}

\begin{figure}[H]
	\centering
	\includegraphics[width=0.95\textwidth]{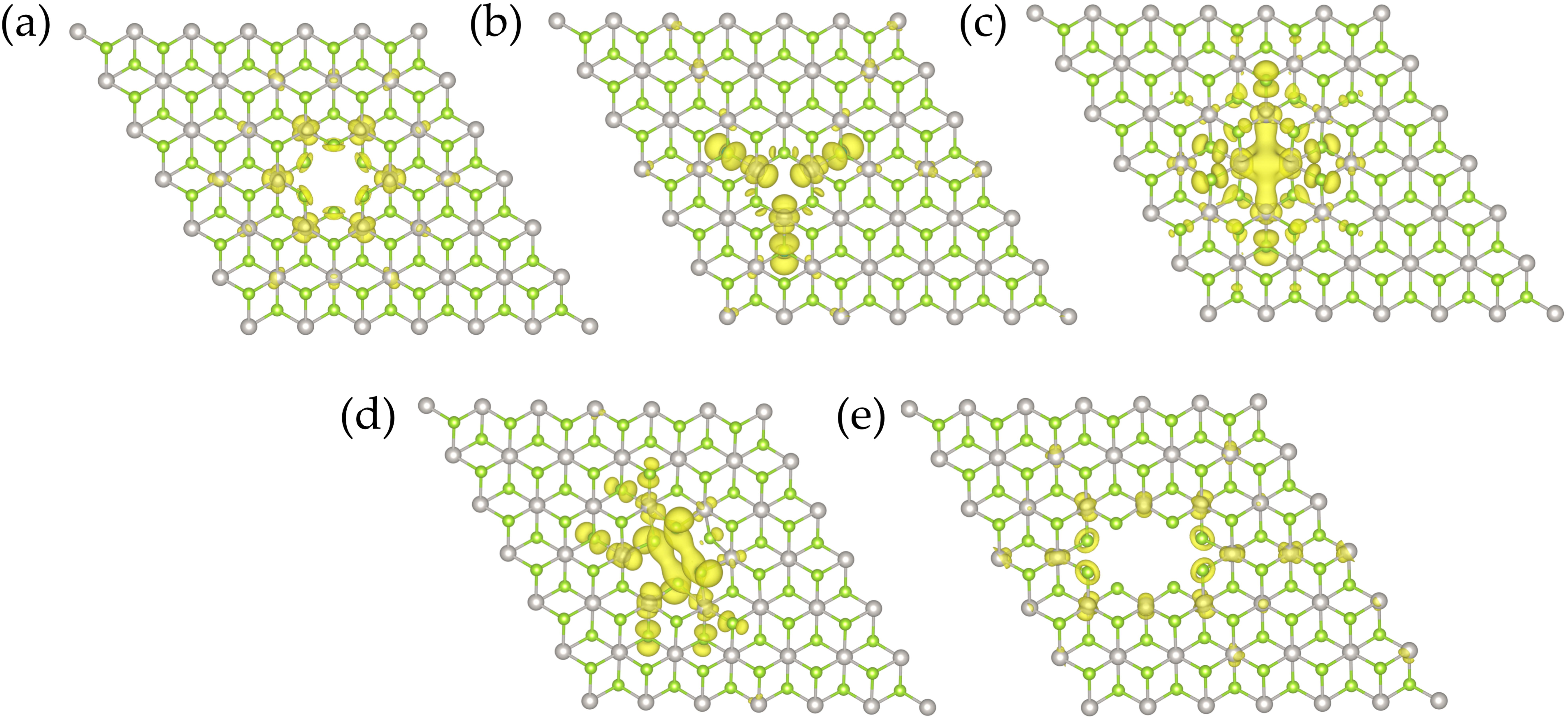}
	\caption{Representative band-decomposed charge densities of selected in-gap defect states for (a) $V_{\mathrm{Pt}}$, (b) $V_{\mathrm{Se}}$, (c) $V_{2\mathrm{Se}}$, (d) $V_{\mathrm{Pt+Se}}$, and (e) $V_{2\mathrm{Pt}+2\mathrm{Se}}$ in monolayer PtSe$_2$.}
	
	\label{fig:figS4}
\end{figure}

\begin{figure}[H]
	\centering
	\includegraphics[width=0.9\textwidth]{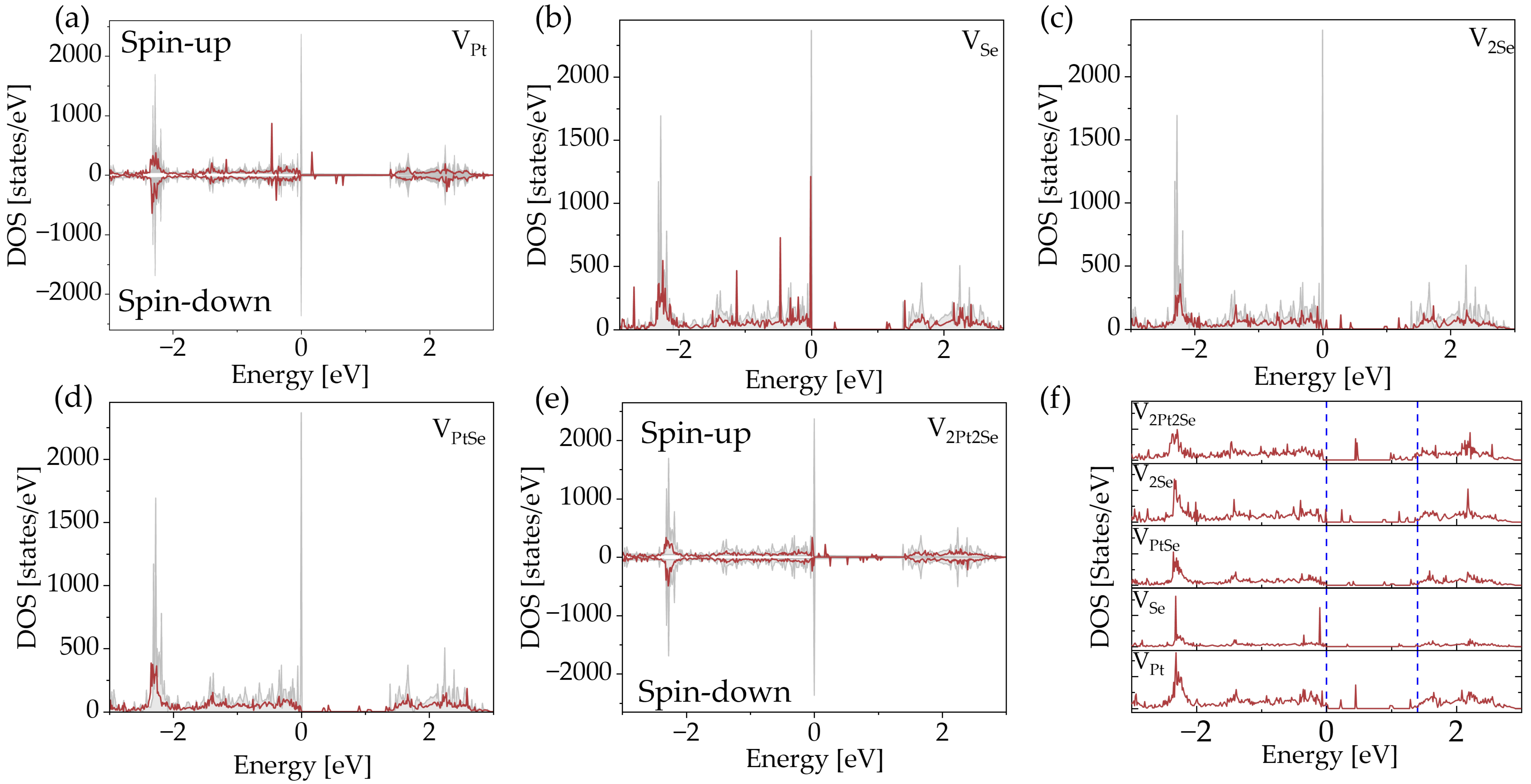}
	\caption{Spin-polarized TDOS of defective PtSe$_2$ calculated without SOC: 
		(a) V$_{\mathrm{Pt}}$, (b) V$_{\mathrm{Se}}$, (c) V$_{\mathrm{2Se}}$, 
		(d) V$_{\mathrm{Pt}+\mathrm{Se}}$, and (e) V$_{2\mathrm{Pt}+2\mathrm{Se}}$.}
	
	\label{fig:figS5}
\end{figure}

\begin{figure}[H]
	\centering
	\includegraphics[width=0.85\textwidth]{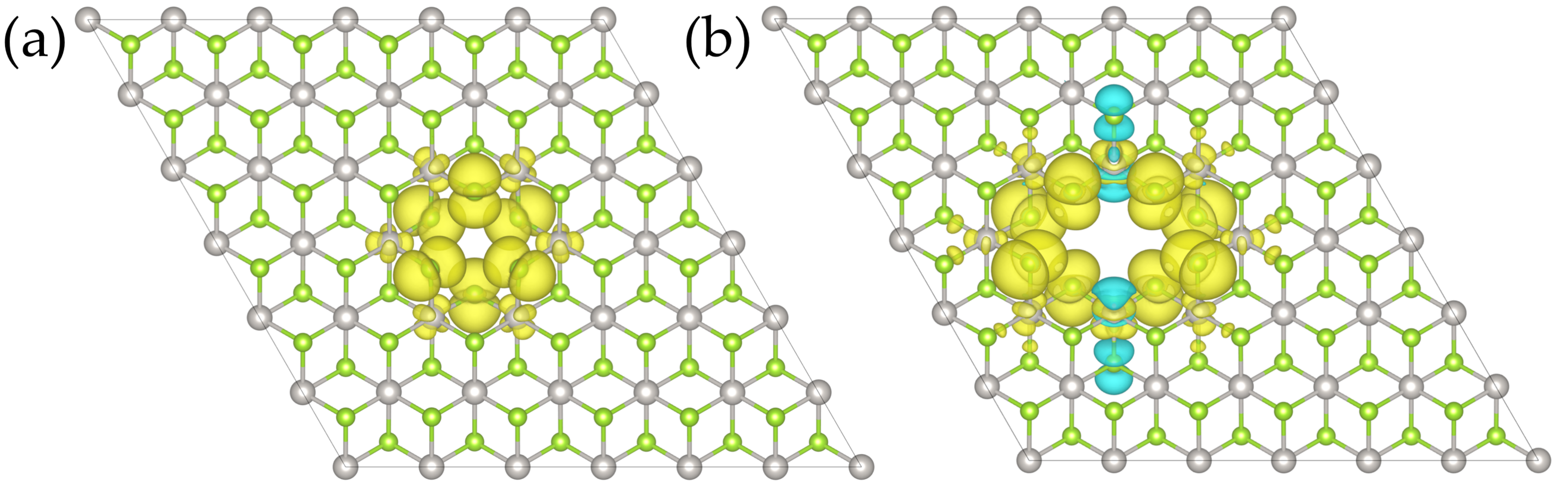}
	\caption{Spin-density isosurface plots of the $V_{\mathrm{Pt}}$(a) and $V_{2\mathrm{Pt}+2\mathrm{Se}}$(b) systems.}
	
	\label{fig:figS6}
\end{figure}

\begin{figure}[H]
	\centering
	\includegraphics[width=0.95\textwidth]{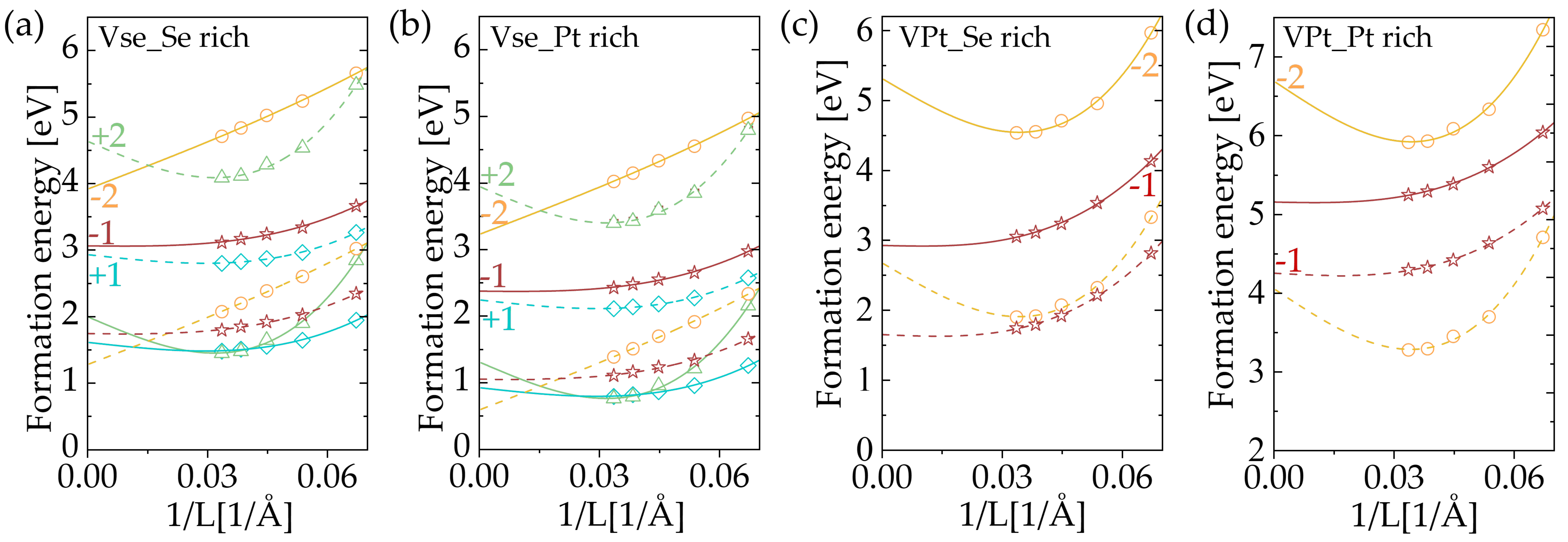}
	\caption{Formation energies of charged vacancies $V_{\mathrm{Se}}$ (a,b) and $V_{\mathrm{Pt}}$ (c,d) in monolayer PtSe$_2$ as a function of inverse system size under Se-rich and Pt-rich conditions. The solid lines represent the formation energies at the valence band maximum, while the dashed lines represent those at the conduction band minimum.}
	\label{fig:figS7}
\end{figure}

\begin{table}[H]
	\centering
	\caption{Finite-size deviations of charged-defect formation energies in monolayer PtSe$_2$. The values are defined as the difference between the formation energy obtained using the \(6\times6\times1\) supercell and the extrapolated dilute-limit value, i.e., \(\Delta E_{\mathrm{FS}}=|E_f(6\times6\times1)-E_f(\infty)\)$|$.}
	\label{TabS2}
	
	\begin{tabular*}{0.8\textwidth}{@{\extracolsep{\fill}}ccccccc}
		\hline\hline
		Charged defect 
		& \(V_{\mathrm{Se}}^{+1}\) 
		& \(V_{\mathrm{Se}}^{+2}\) 
		& \(V_{\mathrm{Se}}^{-1}\) 
		& \(V_{\mathrm{Se}}^{-2}\) 
		& \(V_{\mathrm{Pt}}^{-1}\) 
		& \(V_{\mathrm{Pt}}^{-2}\) \\
		\hline
		\(\Delta E_{\mathrm{FS}}\) (eV) 
		& 0.06 
		& 0.35 
		& 0.18 
		& 1.11 
		& 0.31 
		& 0.61 \\
		\hline\hline
	\end{tabular*}
\end{table}

\begin{figure}[H]
	\centering
	\includegraphics[width=0.95\textwidth]{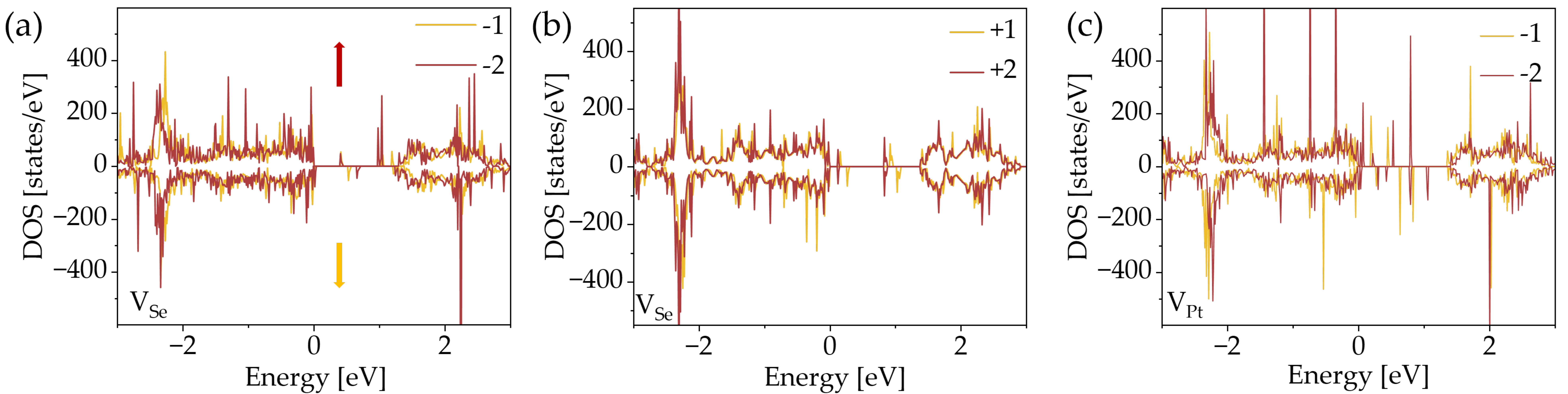}
	\caption{ TDOS for charged vacancies in 2D PtSe$_2$. (a) V$_{\mathrm{Se}}$ in the $-1$ and $-2$ charge states; (b) V$_{\mathrm{Se}}$ in the $+1$ and $+2$ charge states; (c) V$_{\mathrm{Pt}}$ in the $-1$ and $-2$ charge states.}
	\label{fig:figS8}
\end{figure}

\begin{figure}[H]
	\centering
	\includegraphics[width=0.95\textwidth]{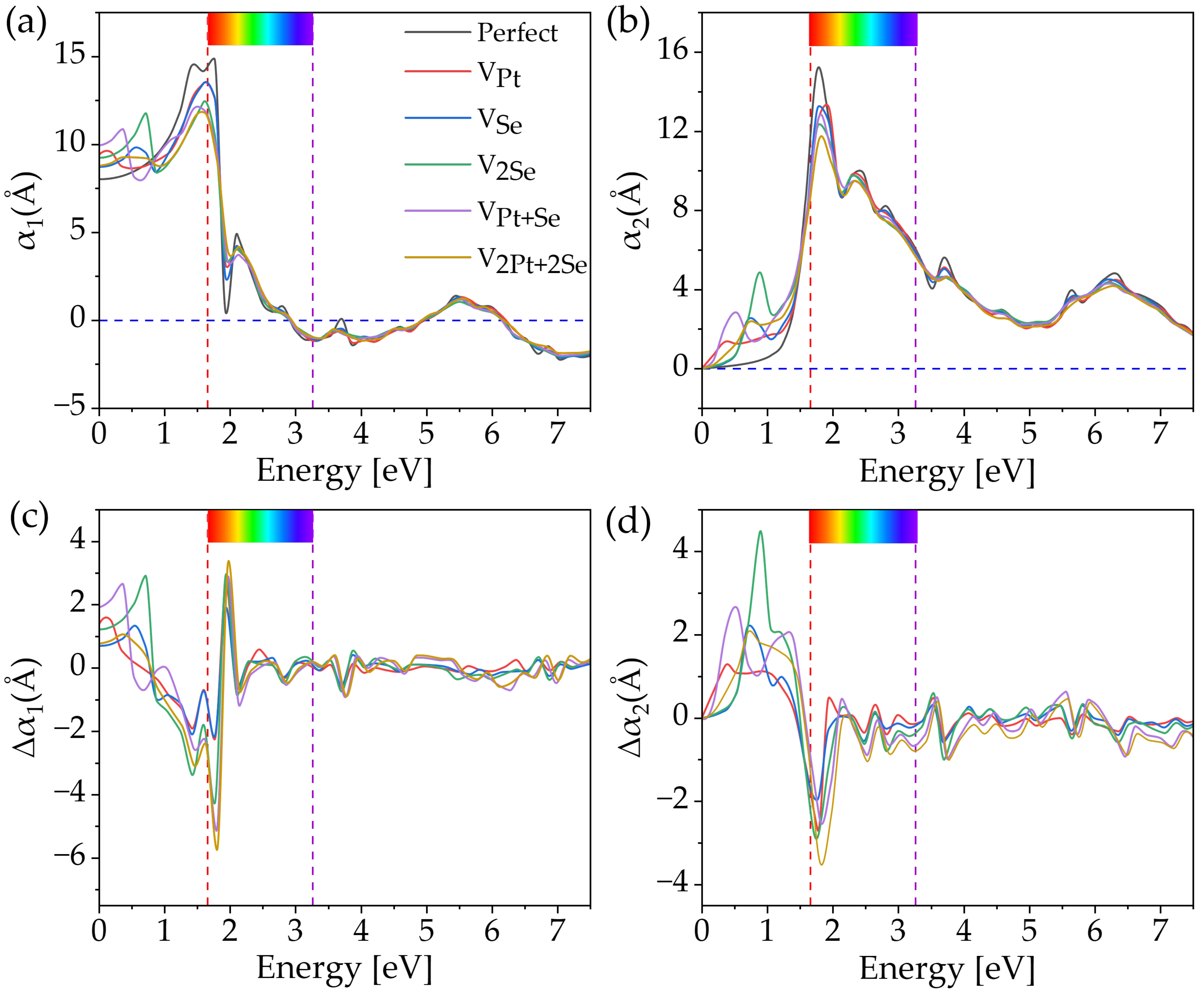}
	\caption{Optical polarizability along the \(y\)-direction of pristine monolayer PtSe$_2$ and its five point-defect structures as a function of photon energy. (a) Real part \(\alpha_1\) and (b) imaginary part \(\alpha_2\) of the polarizability. (c) Differential real part \(\Delta\alpha_1\) and (d) differential imaginary part \(\Delta\alpha_2\) induced by the defects. The colored bar at the top indicates the visible spectral range.}
	\label{fig:figS9}
\end{figure}

For the hydrogen adsorption calculations, one H atom was adsorbed in a $6\times6\times1$ supercell, corresponding to a dilute H-coverage limit. For pristine monolayer PtSe$_2$, representative surface Se and Pt sites were considered. For each defective configuration, H adsorption was examined at representative Se and Pt sites adjacent to the vacancy. After structural relaxation, the hydrogen adsorption energy was calculated as
\begin{equation}
	\Delta E_{H^*}=E_{\mathrm{slab}+H}-E_{\mathrm{slab}}-\frac{1}{2}E_{H_2},
\end{equation}
where $E_{\mathrm{slab}+H}$, $E_{\mathrm{slab}}$, and $E_{H_2}$ are the total energies of the H-adsorbed slab, the corresponding clean slab, and gas-phase H$_2$, respectively. The hydrogen adsorption free energy was then estimated by
\begin{equation}
	\Delta G_{H^*}=\Delta E_{H^*}+\Delta E_{\mathrm{ZPE}}-T\Delta S_H.
\end{equation}
Following the commonly used thermodynamic approximation for HER descriptors, the combined zero-point-energy and entropy correction, $\Delta E_{\mathrm{ZPE}}-T\Delta S_H$, was taken as 0.24 eV at 298 K. The relative exchange current density was estimated using
\begin{equation}
	j_0 \propto \exp\left(-\frac{|\Delta G_{H^*}|}{k_BT}\right),
\end{equation}
which was used only as a qualitative descriptor to compare the relative H-binding tendency of different defect sites. It should be noted that solvent effects, electrode potential, pH dependence, surface charging, and explicit electrochemical interfaces were not included; therefore, the calculated $\Delta G_{H^*}$ values should be interpreted as relative thermodynamic indicators rather than quantitative predictions of HER activity under realistic electrochemical conditions.
The hydrogen adsorption free energy $\Delta G_{H^*}$ is commonly used as a thermodynamic descriptor to evaluate the H-binding tendency of catalyst surfaces. For HER-related adsorption thermodynamics, when \(\Delta G_{H^*}\) is close to 0 eV, it generally indicates a favorable balance between H adsorption and desorption.

\begin{figure}[H]
	\centering
	\includegraphics[width=0.95\textwidth]{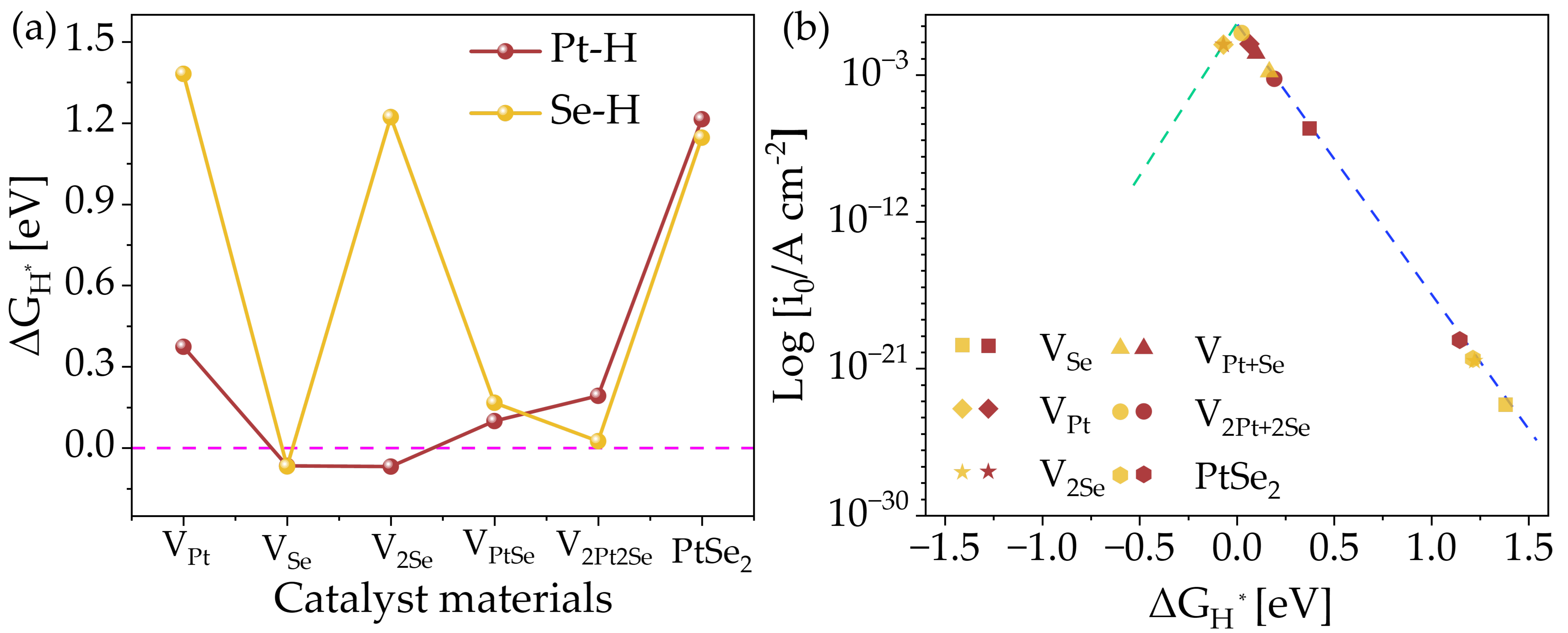}
	\caption{(a) $\Delta G_{\mathrm{H^*}}$ for the hydrogen evolution reaction at Se and Pt sites on monolayer PtSe$_2$ and its various defective structures. The red curves represent H adsorption on Pt atoms, the yellow curves represent H adsorption on Se atoms, and the purple dashed line denotes the reference line for $\Delta G_{\mathrm{H^*}}$. (b) Volcano plot of $\Delta G_{\mathrm{H^*}}$ versus relative exchange current density for monolayer PtSe$_2$ and different defective structures.}
	\label{fig:figS10}
\end{figure}

Fig.~\ref{fig:figS10}(a) shows the $\Delta G_{H^*}$ values for H adsorption at representative Se and Pt sites in pristine monolayer PtSe$_2$ and different defective structures. For pristine monolayer PtSe$_2$, the $\Delta G_{H^*}$ values at both Se and Pt sites deviate markedly from 0 eV, indicating unfavorable H-binding thermodynamics. After introducing vacancy defects, the H adsorption free energy at defect-adjacent sites is significantly modulated, reflecting that vacancy-induced local electronic reconstruction affects the surface adsorption behavior. Among the considered structures, the Se site adjacent to the $V_{2\mathrm{Pt}+2\mathrm{Se}}$ defect exhibits a $\Delta G_{H^*}$ value of 0.025 eV, which is the closest to the thermoneutral value among all sampled sites and closer to 0 eV than the Pt(111) benchmark value of $-0.09$ eV. The volcano plot in Fig. ~\ref{fig:figS10}(b) also shows that this site is located near the apex of the descriptor curve and corresponds to the largest relative exchange-current-density descriptor among all sampled sites.

\nociteSI{REVTEX42Control,apsrev42Control}
\bibliographystyleSI{apsrev4-2}
\bibliographySI{SI}

\end{document}